\documentclass[%
 reprint,amsmath,amssymb,aps]{revtex4-2}
\usepackage{subfigure}
\usepackage{graphicx}
\usepackage{xcolor}
\usepackage{bm}
\usepackage{hyperref}
\usepackage{times}

\begin{document}
\title{Closed ecosystems extract energy through self-organized nutrient cycles}
\author{Akshit Goyal$^1$} 
\email{akshitg@mit.edu}
\author{Avi I. Flamholz$^2$} 
\author{Alexander P. Petroff$^3$} 
\author{Arvind Murugan$^4$}
\email{amurugan@uchicago.edu}
\address{$^1$ Department of Physics, Massachusetts Insitute of Technology, Cambridge, MA 02139.}
\address{$^2$ Division of Biology and Biological Engineering, California Institute of Technology, Pasadena, CA 91125.}
\address{$^3$ Department of Physics, Clark University, Worcester, MA 01610.}
\address{$^4$ Department of Physics, University of Chicago, Chicago, IL 60637.}

\begin{abstract}
Our planet is roughly closed to matter, but open to energy input from the sun. However, to harness this energy, organisms must transform matter from one chemical (`redox') state to another. For example, photosynthetic organisms can capture light energy by carrying out a pair of electron donor and acceptor transformations (e.g., water to oxygen, CO$_2$ to organic carbon).  Closure of ecosystems to matter requires that all such transformations are ultimately balanced, i.e., other organisms must carry out corresponding reverse transformations, resulting in cycles that are coupled to each other. A sustainable closed ecosystem thus requires self-organized cycles of matter, in which every transformation has sufficient thermodynamic favorability to maintain an adequate number of organisms carrying out that process.
Here, we propose a new conceptual model that explains the self-organization and emergent features of closed ecosystems. We study this model with varying levels of metabolic diversity and energy input, finding that several thermodynamic features converge across ecosystems. Specifically, irrespective of their species composition, large and metabolically diverse communities self-organize to extract $\approx10\%$ of the maximum extractable energy, or $\approx100$ fold more than randomized communities. Moreover, distinct communities implement energy extraction in convergent ways, as indicated by strongly correlated fluxes through nutrient cycles. As the driving force from light increases, however, these features – fluxes and total energy extraction – become more variable across communities, indicating that energy limitation imposes tight thermodynamic constraints on collective metabolism. 
\end{abstract}

\maketitle

The biosphere on Earth is open to energy input in the form of sunlight but is closed to matter exchange \cite{odum1971fundamentals,cleveland2008fundamental}. Only some organisms (phototrophs) can directly capture this light energy which then percolates through the biosphere and supports other organisms (heterotrophs). However, redox metabolism links the energy transfer from the sun to the biosphere to the coordinated cycling of matter by phototrophs and heterotrophs \cite{brock2003brock, falkowski2008microbial}. Specifically, phototrophs can capture energy from light only by coupling it to a transformation of matter from one oxidation state to another, e.g., converting CO$_2$ to organic carbon and H$_2$S to sulfates or H$_2$O to oxygen \cite{zagarese2021patterns,odum1956primary}. Similarly, heterotrophs extract chemical energy (e.g., as ATP) by transforming matter, often in the other direction, e.g., organic carbon to CO$_2$, and sulfates to H$_2$S or O$_2$ to H$_2$O \cite{kirchman1994uptake}. 

These two basic aspects of the biosphere --- closure to matter and redox metabolism --- introduce strong constraints that ecosystems must overcome to extract the energy in sunlight and remain out of equilibrium. Organisms performing different transformation of a nutrient must have coordinated activity, e.g., the rate of conversion of nutrients (e.g., CO$_2$ or sulfate) by phototrophs must be balanced by the rate at which heterotrophs recycle their products (e.g., sugars or sulfide) and regenerate them \cite{rillig2019microbial, esteban2015temporal, bush2015redox, christie2017nutrient}. Thus, nutrients must be converted between oxidation states by different species in a cycle, with their fluxes balanced at steady state. In this sense, every photon captured by a closed ecosystem requires moving a specific amount of matter around one of several redox cycles.

It is not clear how easily such cycles can be created and sustained, given all the constraints that ecosystems must obey.  Each conversion must be thermodynamically favorable enough to provide organisms carrying out that conversion with sufficient energy to sustain themselves. Further, nutrient cycles are tightly interconnected \cite{brock2003brock, falkowski2008microbial}. First, the fluxes in different nutrient cycles are linked by the conservation of redox electrons \cite{kracke2015microbial} - the flux of transformations that donate redox electrons (e.g., CO$_2$ to organic carbon \cite{brock2003brock} must be matched by the flux transformations that accept those electrons (e.g., sulfides to sulfates).  Second, each species performs multiple conversions and participates in several nutrient cycles. Thus, changes in the fluxes of one cycle can percolate to others through their effect on species abundances, which may then increase or decrease the flux of the original cycle, and consequently make the simultaneous balance of all cycles challenging \cite{taub2013pressure,matsui2000direct,taub1974closed,obenhuber1988carbon}. Finally, these constraints must be satisfied without help from any central coordinator; in order to extract energy from light \cite{de2021closed, pagaling2017assembly}, closed ecosystems must self-organize the thermodynamics and fluxes of all cycles in a manner consistent with the above constraints.

Despite this challenge, ecosystems have stably re-established nutrient cycles after major perturbations throughout the history of life on Earth, e.g., the oxygenation of the atmosphere \cite{lyons2014rise}, ice ages \cite{imbrie1986ice}, and bolide impacts \cite{kasting1990bolide}. Is the emergence, stability, and resilience of these self-organized non-equilibrium systems surprising? Further, the rules of self-organization are locally greedy --- each species grows if it can extract sufficient energy for itself. How globally efficient we should expect ecosystems to be at extracting energy from light, given that they self-organize through such locally greedy rules?

To answer these questions, we propose and study a new conceptual model of microbial ecosystems, the primary catalysts of nutrient cycles for most of Earth’s history. Prevailing models of microbial ecosystems are not suited to thinking about closed ecosystems, because they insufficiently account for the relevant constraints and coupling. Our model combines the redox view of metabolism with ecological dynamics, allowing us to study the emergence and stability of multiple nutrient cycles in a closed setting. By simulating our model, we find that across a range of environmental conditions, once enough species are added, ecosystems almost always self-organize to a stable state where they can spontaneously recycle multiple nutrients, resulting in energy extraction. Even though distinct ecosystems contain different organisms, the fluxes of their nutrient cycles are very similar (convergent). However, as the driving force from light increases, the fluxes of nutrient cycles become more variable (divergent) and species-dependent. Remarkably, in all these cases, ecosystems are very efficient at extracting energy. The energy extracted from spontaneous cycling is $\approx$10\% of the maximum extractable energy, or $\approx$100 fold more than randomized communities. 

These results advance our understanding of how several coupled nutrient cycles spontaneously emerge and stabilize as a result of many interacting components. They also highlight the efficiency with which ecosystems extract energy from an external source like light. 
Finally, our work establishes closed ecosystems as paradigmatic examples of systems that self-organize to determine their displacement from equilibrium, in contrast to traditionally studied non-equilibrium systems in physics where the displacement from equilibrium is fixed \cite{qian2007phosphorylation,hill2013free}.

\section*{Results}
\noindent
\textsf{\textbf{An ecological model of thermodynamically-constrained nutrient cycles.}} Our model describes a self-sustaining ecosystem in which $S$ microbial species collectively recycle a set of environmental resources through sets of $R$ thermodynamically-constrained redox transformations. Each species in the ecosystem corresponds to a different metabolic type, depending on the subset of these $R$ transformations it can perform to maintain itself. Individuals of each species $\alpha$ extract an energy flux $\mathcal{E}^\alpha_{\text{tot}}$ depending on the net energy released by coupling transformations, growing if they extract more than a prescribed maintenance energy $\mathcal{E}_{\text{maint}}$, and dying if they extract less. Species dynamics modify the resource concentrations, making each transformation more or less thermodynamically favorable, and consequently changing the energy extracted by individuals of each species. Eventually, the ecosystem self-organizes to a steady state characterized by two sets of emergent quantities: (1) the abundances of each surviving species $N_\alpha$, where each individual extracts exactly $\mathcal{E}_{\text{maint}}$, and (2) the fluxes of each of the $R$ resource transformation cycles. 

The key ingredient in our model --- which distinguishes it from conventional ``consumer-resource'' models of ecosystems \cite{goyal2018diversity,marsland2019available,goldford2018emergent,tikhonov2017collective,macarthur1970species,posfai2017metabolic} --- is that resources are not single molecules with a certain energy content (Fig. 1a), but instead redox transformations (half-reactions) whose energy content is determined by electron and thermodynamic constraints (Fig. 1b). All $R$ resources correspond to transformations $O_i \leftrightarrow R_i$ between pairs of molecules $O_i,R_i$, representing the oxidized and reduced forms. A redox tower orders all resource pairs $(O_i, R_i)$ by their chemical potential $\mu_i$, from least energetically favorable $O_i \to R_i$ conversion (top) to most (bottom) (Fig. 1a). These chemical potentials $\mu_i$ are given by a standard state potential $\mu_i^{0}$ \cite{brock2003brock, jin2003new} and an adjustment due to concentrations of , i.e., $\mu_i = \mu_i^{0} - \log O_i/R_i$.
Each species $\alpha$ may exploit a specific subset of these transformations specified by kinetic coefficients $e_{i\alpha} \geq 0$; e.g., if $e_{i\alpha} \neq 0$, species $\alpha$ can transform $R_i \to O_i$ with kinetic coefficient $e_{i\alpha}$, releasing electrons at a potential $\mu_i$ which are then absorbed by another transformation $O_j \to R_j$ that the species participates in. The potential difference drop experienced by the electron is the energy available to this species. In practice, the electrons are transferred by an electron carrier (e.g., NADH) in the cell that is at potential $\mu_{\text{carrier},\alpha}$ intermediate to $\mu_i$ and $\mu_j$. We will assume there is only one electron carrier pool common to all redox transformations in species $\alpha$.

Without any external energy source, the chemical potentials $\mu_i$ in the redox tower will obey detailed balance \cite{hill2013free, qian2007phosphorylation}; i.e., as an electron is transferred in a loop through a cycle of redox transformations by different species, the net change in the electron's energy must be zero. Consequently, in such a closed system, some species will have to \emph{provide} energy to move the electron `uphill' during metabolism, rendering such ecosystems not viable. 

\begin{figure*}[ht!]
    \centering
    \includegraphics[width=0.85\textwidth]{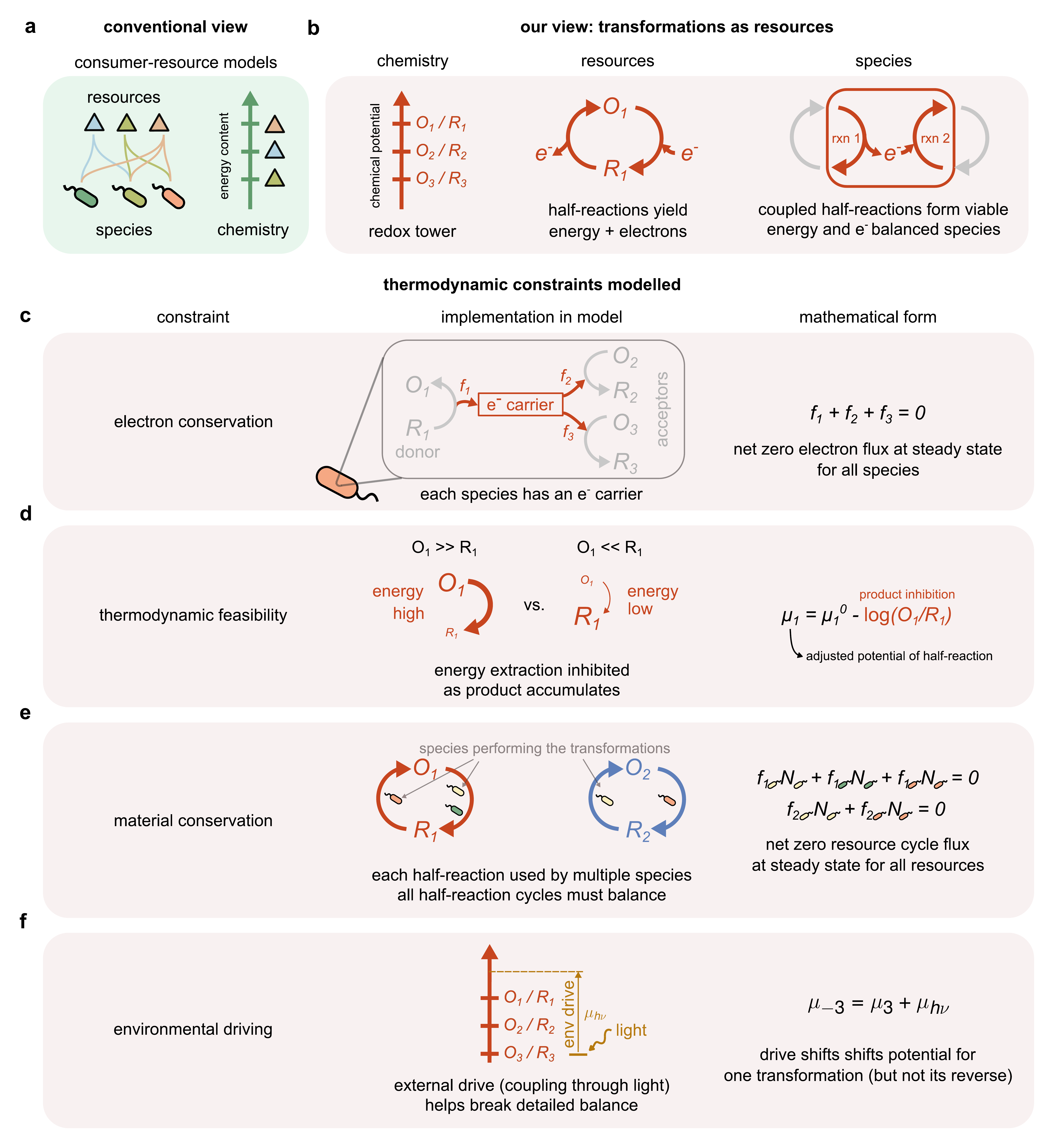}
    \caption{\scriptsize\textsf{\textbf{An ecological model of thermodynamically-constrained nutrient cycles.} (a) In conventional models, species 
    consume different molecules (triangles) that serve as resources with energy content. 
    (b) In our model, resources are \emph{transformations} of molecules, e.g., from an oxidized form $O_i$ to their reduced form $R_i$. Species gain energy by coupling transformations (or redox half-reactions) that donate electrons with transformations that accept electrons. 
    Each (microbial) species is defined by the transformations it can carry out. 
    Transformation fluxes must satisfy thermodynamic and conservation constraints in the ecosystem's non-equilibrium steady state:  
     (c) Electron conservation: The fluxes $f_i^\alpha$ in different half-reactions carried out by a species $\alpha$ must sum to zero, i.e., $\sum_i f_i^\alpha = 0$. We assume that each microbial species has one electron carrier that shuttles electrons between the half-reactions.  (d) Thermodynamic feasibility: The thermodynamic driving force (chemical potential) of each transformation, and consequently energy available, is calculated using thermodynamic principles, accounting for inhibition due to the concentrations of $O_i$ and $R_i$. 
     (e) Material conservation: 
     Fluxes in different parts of each molecular cycle (shown in red and blue), summed over contributions from all species, must be balanced. 
     (f) Detailed balance breaking:  
     Energy available by transforming matter in cycles is constrained by breaking of detailed balance. Detailed balance is broken by coupling some transformations (here, $O_3 \to R_3$) to an energy source (e.g., light) which shifts their potential by $\mu_{h\nu}$ but does not affect the reverse transformation ($R_3 \to O_3$).  
     }}
    \label{fig:1}
\end{figure*}

We will assume that detailed balance is broken across the redox tower because some transformations, say $R_j \to O_j$ are  coupled to an external energy (but not matter) source (e.g., coupling the transformation H$_2$O $\to$ O$_2$ to sunlight during photosynthesis). Consequently, the chemical potential of $R_j \to O_j$ is shifted $\mu_{j} = \mu_{-j} + \mu_{h \nu}$ where $\mu_{j}$ is the chemical potential for the reverse transformation $O_j \to R_j$ (not coupled to light); the term $\mu_{h \nu}$ breaks detailed balance and is a key feature of the physical environment whose impact on ecosystems we will explore below.  



%
The redox formalism allows us to correctly model the thermodynamic constraints on collective energy extraction. To maintain themselves at steady state, species in our model ecosystems must satisfy three constraints: electron conservation, material conservation and sufficient energy extraction. To conserve electrons, the total flux of electron donor and acceptor transformations must balance within each species  (Fig. 1c, Methods). The per capita flux in transformation $O_i \to R_i$ due to species $\alpha$ is determined by species abundance $N_\alpha$, molecular concentration $O_i$, and by thermodynamic driving forces $\Delta \mu_{i \alpha}$, 
\begin{equation}
 f_{i \alpha} = \frac{ \cdot O_i }{ ( K_N + \sum_{\beta=1}^{S} \frac{N_\beta}{k_{i\beta}} )},
\end{equation}
with $k_{i\alpha}$ determined by thermodynamic forces and fluxes (derivation in Supplementary Text),
\begin{equation}
 k_{i \alpha} = e_{i\alpha} \cdot (1 - e^{-\Delta \mu_{i\alpha}}) 
\end{equation}
where $\Delta \mu_{i\alpha} $ is the change in potential of an electron released by transformation $O_i \to R_i$ and transferred to the electron carrier. Hence $\Delta \mu_{i\alpha} = \mu_{\text{\text{carrier}}, \alpha} - \mu_{i}^0 + \log(O_i/R_i)$ where $\mu_i^0$ is the standard state chemical potential of the transformation $O_i \to R_i$ and $\mu_{\text{\text{carrier}}, \alpha}$ is the chemical potential of the electron carrier for in species $\alpha$. 

These fluxes $f_{i \alpha}$ change the concentrations of $O_i, R_i$ through the following dynamics:
\begin{equation}
    \frac{dO_i}{dt} = -\sum_{\alpha=1}^{S} f_{i\alpha} N_\alpha + \sum_{\beta=1}^{S} f_{i\beta} N_\beta 
\end{equation}
where $f_{i\alpha}$ is the flux of the transformation $O_i \rightarrow R_i$ performed by an individual of species $\alpha$ (as given in equation (1)), and $f_{i\beta}$ is the flux of the transformation $R_i \to O_i$ (similar to equation (1), but proportional to the reactant concentration $R_i$, not $O_i$ by individuals of species $\beta$. The first sum goes over all species transforming $O_i \to R_i$ and the second sum over species capable of the reverse. Similar equations hold for $R_i$.

Each species $\alpha$ extracts a per capita energy flux $\mathcal{E}_{\text{tot}}^{\alpha}$ by coupling electrons between transformations at different potentials:
\begin{eqnarray}
     \mathcal{E}_{\text{tot}}^{\alpha} &=& \sum_{i=1}^{2R} \mathcal{E}_{i\alpha} = 
     \sum_{i=1}^{2R} f_{i\alpha}  \cdot \Delta \mu_{i\alpha}. 
\end{eqnarray}
A species grows in abundance if this extracted energy exceeds a prescribed per capita maintenance energy $\mathcal{E}_{\text{maint}}$:

\begin{equation}
    \frac{1}{N_\alpha} \frac{dN_\alpha}{dt} = \mathcal{E}_{\text{tot}}^{\alpha} - \mathcal{E}_{\text{maint}}.
\end{equation}

%

Finally, to conserve materials, as species in the ecosystem couple different half-reactions and transform resources from one form to another, all resource cycles must be balanced (Fig. 1e). Together, these constraints can be summarized as:

\begin{align}
    \underline{\text{electron conservation:}} && \sum_{i=1}^{2R} f_{i\alpha} &= 0,\\
    \underline{\text{energy balance:}} && \sum_{i=1}^{2R} f_{i\alpha} \cdot \Delta \mu_{i\alpha} &= \mathcal{E}_{\text{maint}},\\
    \underline{\text{material conservation:}} && \sum_{\alpha=1}^{S} N_\alpha f_{i\alpha} &= 0.
\end{align}

The last equation amounts to assuming that the ecosystem is fully closed to matter, and open only to an external source of energy (here, light energy $\mu_{h\nu}$) that breaks detailed balance for the chemical potentials $\mu_i$. While we use a fully materially closed ecosystem as an extreme case to illustrate our model (Figs. S1 and S2), many of our results hold for partially closed ecosystems as well, where some of the resources can be exchanged with the environment and equation (6) is modified (Fig. S3). 

In addition to providing energy through transformations, matter also directly contributes to biomass \cite{scott2010interdependence}. We assume that the amount of matter sequestered as biomass is insignificant compared to the total availability ($O_i + R_i$) and only focus on energy in this work. We leave an analysis of the dual role of matter in providing energy (through transformations) and biomass to future work.

\begin{figure*}[ht!]
    \centering
    \includegraphics[width=0.9\textwidth]{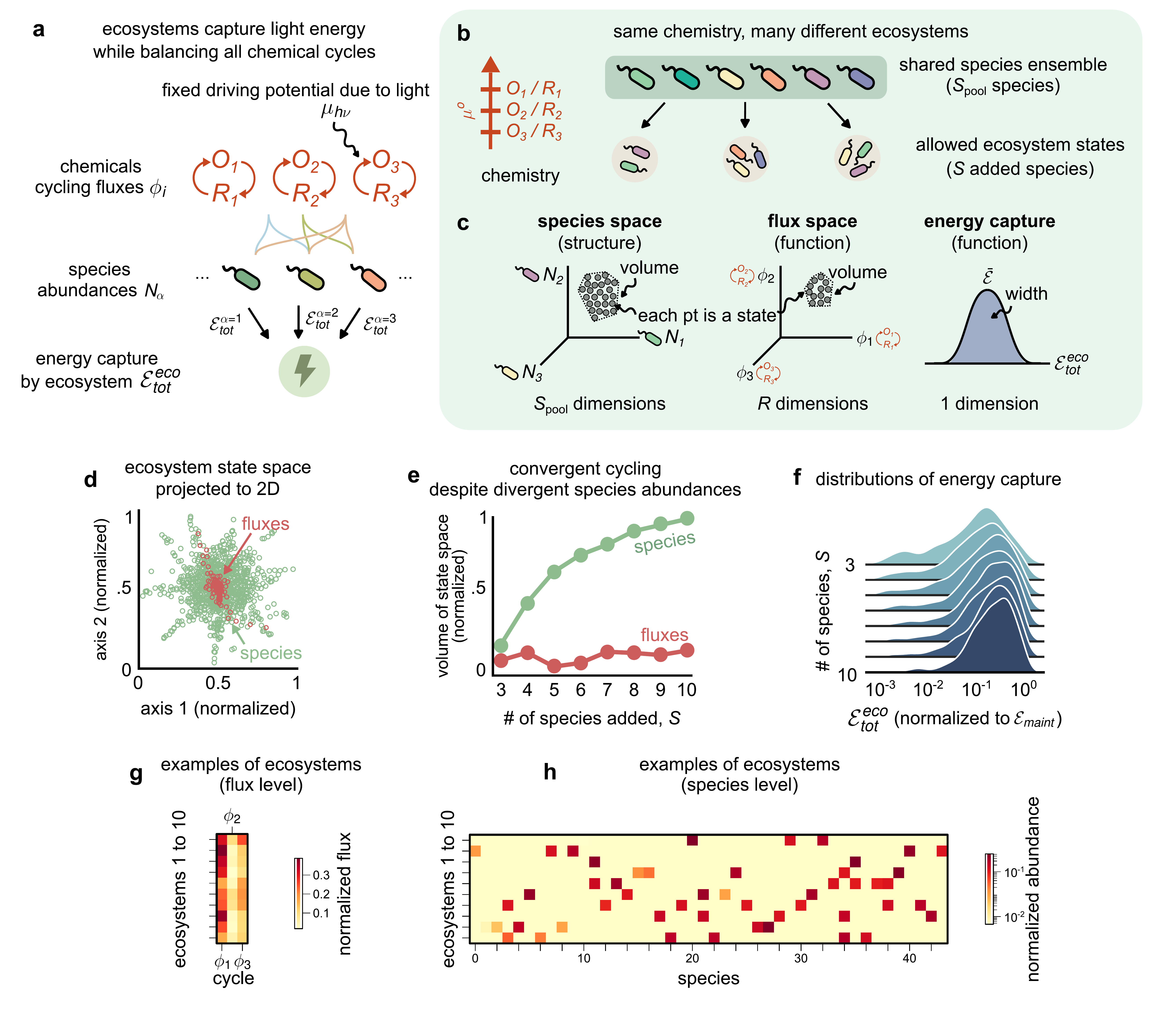}
    \caption{\scriptsize\textsf{\textbf{Ecosystems spontaneously implement correlated nutrient cycles and convergent energy extraction while containing distinct species.} (a) Schematic showing how ecosystems in our model extract energy from externally supplied light (with driving potential $\mu_{h\nu}$), which affects the redox potentials of certain half-reactions. At steady state, all resources $(O_i, R_i)$ are cycled with fluxes $\phi_i$. Each microbial species (colored ellipses with wiggles) with abundances $N_\alpha$ carries out a subset of half-reactions (undirected colored links). By performing metabolic transformations, each surviving individual must extract a maintenance energy $\mathcal{E}_{\text{maint}}^\alpha$;  collectively, the entire ecosystem extracts an energy flux  $\mathcal{E}_{\text{tot}}^{\text{eco}}$. (b) Schematic showing a given physical environment consisting of $R=3$ redox transformations (resources), and a pool of $S_{\text{pool}}=100$ species (top rectangle) used to randomly assemble constraint-satisfying ecosystems in simulations. (c) Cartoon showing possible ecosystem solutions in a space of species abundances (left), resource cycle fluxes (middle), and a distribution of collective energy extracted ($\mathcal{E}_{\text{tot}}^{\text{eco}}$; right). (d) Scatter plot showing the space of species (green) and fluxes (red) from 1,000 randomly assembled ecosystems from simulation, projected to two dimensions using multi dimensional scaling (MDS) (Methods). (e) Line plot showing how the volume of the species (green) and flux (red) spaces scales with the number of species added, $S$, in assembled ecosystems. The flux space volume grows much slower than species space volume, indicating convergence in the function (fluxes) of constraint-satisfying ecosystems. (f) Distributions of the total energy extracted by ecosystems, $\mathcal{E}_{\text{tot}}^{\text{eco}}$, as a function of $S$. As ecosystems become more species-rich, ecosystems extract greater average energy with greater convergence (lower variance). (g--h) Heatmaps showing examples from 10 of the 1,000 randomly assembled ecosystems in (d--f), showing the (g) fluxes and (h) species abundances in detail. Each row shows an ecosystem, while each column shows a resource (in (g)) and a species (in (h)). 
    }}
    \label{fig:2}
\end{figure*}

The constraints encoded in our model naturally give rise to multiple solutions --- there is a large space of ecosystems that satisfy them. To sample this space, we seeded a chosen physical environment with random mixtures of species with random metabolic strategies, and numerically evolved the dynamical equations until a subset of species found a stable composition or went extinct. Each ecosystem was provided with the same `physical environment', i.e., concentrations of $R_i,O_i$ and energy input $\mu_{h\nu}$, but with a different initial set of $S$ species chosen randomly and allowed to reach steady state (Methods).  By simulating 1,000 such ecosystems, we sampled a large space of self-organized ecosystems.

\vskip 10pt
\noindent
\textsf{\textbf{Emergent similarity of nutrient cycles.}} To quantify the size of this space of ecosystems, we measured their structure (set of species abundances) and function (set of resource fluxes $\phi_i$) and total energy extracted $\mathcal{E}_{\text{tot}}^{\text{eco}}$) (Fig. 2c). Each point in structure space represents the  abundances of each of the $S_{\text{pool}}=100$ species in the pool in one of our simulated ecosystems (one species per dimension). Similarly, each point in flux space represents the fluxes in each of the $R=3$ resource cycles in that ecosystem, and each point in the distribution of energy extracted represents the total energy $\mathcal{E}_{\text{tot}}^{\text{eco}}$ captured from the external energy source by the ecosystem. To compare the size of the species and flux spaces, we projected them both onto a common two-dimensional space while preserving pairwise distances between ecosystems and computed the resulting area (see Methods); such a projection allows for a fair comparison of the convergence in species abundances and fluxes which are of different dimensionalities. 
We find that ecosystem-wide fluxes show less variability than the structure of species enabling these fluxes (Fig. 2d); further, while the variability in species abundances rapidly grows with the number of added species, the flux variability does not (Fig. 2e and Fig. S8).
While alternative methods of comparing flux and species variability can change their absolute numbers, we expect this trend of increasing functional convergence with species diversity to be robust.

\begin{figure*}[ht!]
    \centering
    \includegraphics[width=0.7\textwidth]{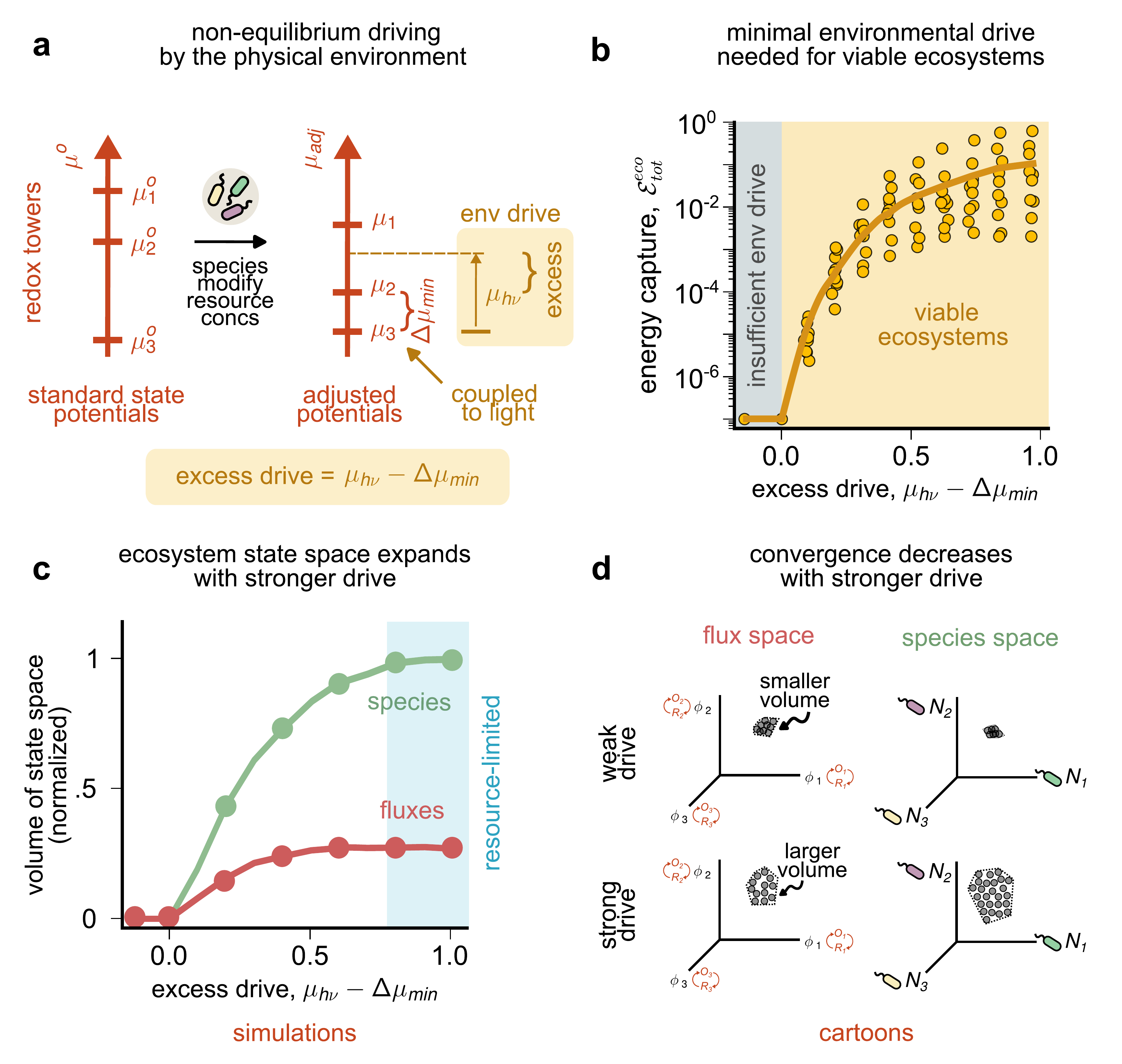}
    \caption{\scriptsize\textsf{\textbf{Nutrient cycles and energy extraction become more variable with stronger environmental driving.} (a) Schematic showing non-equilibrium driving by the environment, modeled as the coupling of one of the half-transformations (at the bottom of the redox tower) to light energy. Environmental driving shifts the chemical potential of of the transformation $R_3 \to O_3$ by $\mu_{h\nu}$, but not the reverse, thus breaking detailed balance. For ecosystems to be viable and satisfy all the modeled thermodynamic constraints, the environmental drive $\mu_{h\nu}$ must exceed the smallest adjusted potential difference $\Delta \mu_{\text{min}}$ between $O_3/R_3$ and the closest redox pair. This minimal drive requirement is a statement about adjusted potentials on the redox tower, not the standard state potentials. The former depend on the species present and their abundances, and are thus self-organized. We quantify $\mu_{h\nu} - \Delta \mu_{\text{min}}$ as the extent of environmental drive beyond the minimal ($\Delta \mu_{\text{min}}$) needed for a viable ecosystem. (b) Scatter plot showing the energy extracted by ecosystems from our model as a function of excess drive. Each point represents a random self-organized ecosystem simulated using our model, under the same conditions as in Fig. \ref{fig:2}, at varying $\mu_{h\nu}$. The solid line shows a moving average. No ecosystems are viable without sufficient environmental driving (gray region). Increasing $\mu_{h \nu}$ increases energy extraction on average (mustard region). (c) Line plot showing the volume of ecosystem state space in species abundances (green) and resource fluxes (red), as in Fig. \ref{fig:2}e, but with varying excess drive $\mu_{h\nu} - \Delta\mu_{\text{min}}$. Ecosystems near equilibrium are strongly constrained; the volume of viable ecosystems grows rapidly with stronger environmental driving, eventually saturating at large $\mu_{h \nu}$. Results are shown from simulations with 1,000 randomly assembled ecosystems. (d) Cartoons showing how flux (red) and species (green) spaces expand (decreasing convergence) with stronger drive, with spaces represented as in Fig. 2c.}}
    \label{fig:3}
\end{figure*}

A nearly constant flux variability, despite increasing species variability, suggests that fluxes are a convergent feature of ecosystems that perform thermodynamically-constrained transformations. 
Indeed, the convergence of fluxes despite species divergence is evident from a few examples of our model ecosystems (Fig. 2g--h, fluxes and species respectively). 
We also found that ecosystems had more convergent structure when coarse-grained by metabolic type (the set of transformations each species could perform) than at the species level (Fig. S4), similar to what has been observed in metagenomic data from field surveys of microbial communities \cite{louca2016high, louca2018function, goldford2018emergent}. 


While functional convergence of fluxes and metabolic types has been noted in other contexts, both in experiments\cite{estrela2022functional, louca2016high} and from other models\cite{george2022functional, fant2021eco}, our model allows us to go further and study convergence of explicitly thermodynamic functions of ecosystems. We find that the distributions of total energy extracted by our model ecosystems also converges, with the mean energy extracted increasing and variance in energy extracted decreasing with diversity, before ultimately saturating (Fig. 2f). 
Finally, we determined the energy contribution $\mathcal{E}_{ij}$ of each pair $(i,j)$ of redox transformations to the total energy harvested; we found that such global `redox strategies' of ecosystems became progressively similar with increasing diversity (Fig. S5).


Thus, we find that thermodynamic constraints of coupled transformations self-organize ecosystems such that collective functions -- here, the distribution of fluxes and energy extracted across cycles -- are similar across distinct ecosystems that differ widely in their structure (species content). 


\vskip 10pt
\noindent
\textsf{\textbf{Collective functions become more variable with stronger environmental driving.}}
The influence of the physical environment (e.g., nutrients supplied) is at the core of all ecology; but the impact of thermodynamic properties of the environment on ecosystem organization has been studied only in models of specific systems (e.g., communities with hydrogenotrophic methanogens where end-product inhibition is crucial \cite{lynch2019modelling, delattre2020thermodynamic, hoh1996practical}). 

In our model, the physical environment consists of a redox tower with standard-state chemical potentials $\mu^0_i$ of the redox transformations $R_i \leftrightarrow O_i$ and the total amounts of matter $R_i + O_i$ available for each transformation. Most critically, the environment also includes an external energy drive such as light that shifts the potential of one redox transformation by $\mu_{h\nu}$ (but does not affect the reverse transformation). 

Without such an external drive $\mu_{h \nu}$, transformations within an ecosystem are constrained to satisfy detailed balance, a defining property of equilibrium systems, and thus no net energy can be extracted\cite{hill2013free, qian2007phosphorylation}.
At the same time, non-equilibrium driving by $\mu_{h\nu}$ does not by itself set the energy extracted by the ecosystem; species abundances must self-organize and ultimately determine the energy extracted.  
Fixing the environmental driving potential $\mu_{h \nu}$ is distinct from fixing the flux of an external resource which has been the focus of previous modeling approaches \cite{marsland2019available, george2022functional, cook2021thermodynamic}. In an electrical circuit analogy, $\mu_{h\nu}$ sets the external voltage while prior approaches typically set an external current. These two approaches are especially distinct in self-organized ecosystems where the analog of `resistance' is not fixed since species abundances evolve through population dynamics.

We sought to understand how external drive $\mu_{h \nu}$ in the physical environment affects the number of viable ecosystems, and their functional convergence. We simulated an ensemble of ecosystems for each of several environmental driving potentials $\mu_{h \nu}$. 
We found that ecosystems can sustain themselves only if  $\mu_{h \nu}$ exceeds the smallest potential difference $\Delta\mu_{\text{min}}$ (Fig. 3b) between all  potentials $\mu_i$ on the redox tower (Fig. 3a). Below this level, ecosystems were energy deprived due to insufficient environmental driving (Fig. 3b, gray region) and not viable. 




Note that the minimal environmental drive $\Delta \mu_{\text{min}}$  for ecosystem viability depends on redox potentials $\mu_i$ that account for product inhibition in the ecosystem self-organized by the species present and not on standard state potentials $\mu^0_i$. 
Thus, in distinction to prior work, our model can determine the minimal environmental drive for ecosystem viability as a function of which species are present (Fig. S6). 

For strong enough driving $\mu_{h \nu}> \Delta \mu_{\text{min}}$, ecosystems self-organized to extract greater collective energy $\mathcal{E}_{\text{tot}}^{\text{eco}}$ with increasing external chemical potential $\mu_{h\nu}$ (Fig. 2b, mustard region). The mean $\mathcal{\bar{E}}_{\text{tot}}^{\text{eco}}$ eventually saturated at large $h\nu$, due to limitation from total resource concentrations $\sum_i (O_i + R_i)$ (Fig. S7). 

Notably, the variance in $\mathcal{E}_{\text{tot}}^{\text{eco}}$ also increased with $\mu_{h\nu}$ (Fig. 3b, yellow circles), suggesting that stronger environmental driving decreased convergence in the collective energy extracted $\mathcal{E}_{\text{tot}}^{\text{eco}}$. In addition, both the species and flux spaces of constraint-satisfying ecosystems decreased in size with decreased environmental driving (Fig. 3c, species volume in green, flux volume in red). The species and flux spaces, as well as collective energy extracted, are emergent features, representing how ecosystems self-organize once the physical environment drives them sufficiently far from equilibrium. Together, these results suggest a relationship between non-equilibrium environmental driving and the degree of convergence in ecosystems. When ecosystems operate in environments that are closer to equilibrium (small $\mu_{h\nu}$), there is a smaller space of distinct ways to cycle all resources successfully and provide maintenance energy for all organisms. Consequently, convergence is strongest in environments that are near equilibrium.
\begin{figure*}[ht!]
    \centering
    \includegraphics[width=0.9\textwidth]{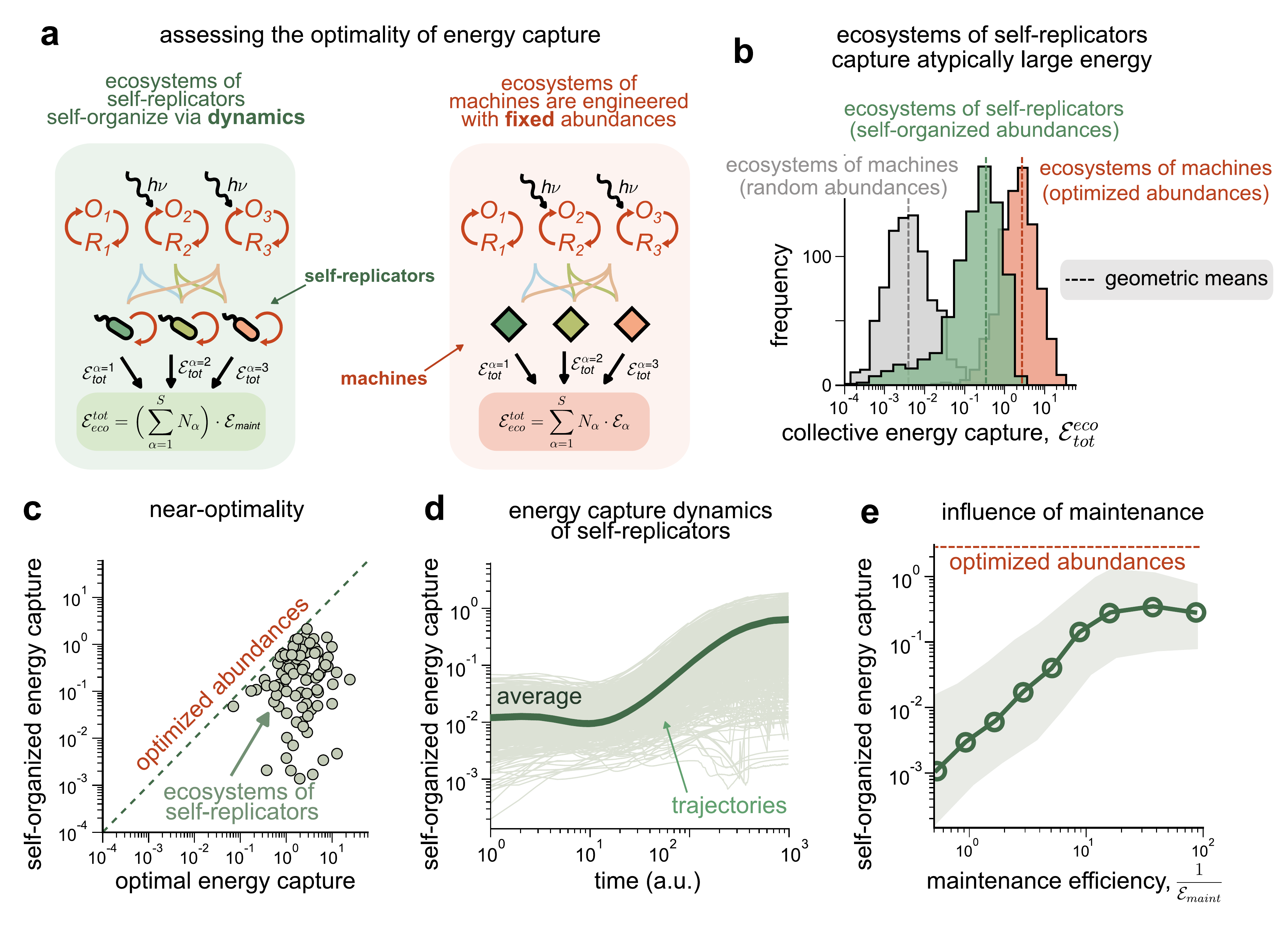}
    \caption{\scriptsize\textsf{\textbf{Ecosystems self-organize to extract a near optimal amount of energy.} 
    (a) We compare energy extracted by ecosystems of living organisms (self-replicators) to a ecosystem of machines with identical metabolic capabilities. Like living organisms, machines come in multiple species with distinct metabolic types and are subject to the same thermodynamic and conservation constraints. However, machines have fixed abundances, not subject to birth-death dynamics driven by maintenance energy requirements found in living systems. 
    (b) Histograms of the total energy extracted $\mathcal{E}_{\text{tot}}^{\text{eco}}$ by ecosystems of machines with random abundances (gray; representing initial conditions of ecosystem assembly), ecosystems of self-replicators with abundances self-organized by birth-death dynamics based on maintenance energy (green), and machines with abundances chosen to maximize $\mathcal{E}_{\text{tot}}^{\text{eco}}$ (orange). Ecosystems of self-replicators reach steady states that extract $\sim 100\times$ more energy than initial random abundances. (c) Scatter plot showing the energy extracted $\mathcal{E}_{\text{tot}}^{\text{eco}}$ by ecosystems with self-organized abundances ($y$-axis) and optimized abundances ($x$-axis); both systems are constituted from the same pool of metabolic strategies. 
    (d) Trajectories showing the dynamics of ecosystems of self-replicators reaching steady states over time (1,000 simulations). 
    (e) Line plot showing how the average $\mathcal{E}_{\text{tot}}^{\text{eco}}$ (solid green) changes as a function of the maintenance efficiency per cell $\mathcal{E}_{\text{maint}}^{-1}$. Each point represents the average collective energy extracted by ecosystems of self-replicators at the specified $\mathcal{E}_{\text{maint}}$ simulated using our model (averaged over 1,000 simulations); the green envelope their s.e.m.; the red dashed line shows the geometric mean of the energy extracted by ecosystems of machines with optimized abundances.  
    }}
    \label{fig:4}
\end{figure*}

\vskip 10pt
\noindent
\textsf{\textbf{Near-optimal energy extraction by self-organized ecosystems.}} The collective energy extracted $\mathcal{E}_{\text{tot}}^{\text{eco}}$, an emergent community function, is not explicitly optimized by any aspect of our model; our population dynamics equations model `greedy' or `selfish' biological species that grow in abundance to extract the most energy that each species can extract in the context of a self-sustaining ecosystem.  

A natural question is how this energy extracted by an ecosystem of locally selfish replicators compares to an alternative community of `metabolic machines' agents whose abundances are determined by global energy optimization but that are still subject to the same thermodynamic and flux constraints. We explored one such alternative framework based on communities of non-living `metabolic machines' identical to self-replicating living species in terms of metabolism but not subject to birth-death population dynamics. More explicitly, each machine species $\alpha$ was identical to living species $\alpha$ in terms of its metabolic properties (e.g., $e_i^\alpha$ which dictate which redox transformations $i$ species $\alpha$ can perform). Further, energy extraction by the machines were subject to all the thermodynamic, electron and material conservation constraints discussed earlier.

We initialized a community of machines with initial random abundances and adjusted those abundances minimally to obey electron and mass conservation constraints (see Methods). We computed the energy extracted by these  communities of effectively random abundances (Fig. 4b, gray). We then evolved these random initial set of abundances in two distinct ways: (1) Global community-wide energy optimization for machines, (2) Local maintenance-energy based population dynamics for self-replicators.

(1) Global optimization for machines: we optimized the abundances $N_\alpha$ of each species $\alpha$, subject to electron and flux constraints to maximize community-wide energy extraction; the machines were not subject to any population dynamics (Methods and Fig. 4a). Consequently, a machine species of type $\alpha$ could exist at any abundance, dictated only by what was optimal for the community as a whole, even if its own energy extracted would have lead to higher or lower abundance (or even extinction) according to `selfish' population dynamics (equation (6)). The energy, as expected, is dramatically higher after such global optimization (Fig. 4b, orange).

(2) Local population dynamics for self-replicators: Starting from random initial abundances, we also ran population dynamics based on maintenance energy (equation (7)). Unlike global optimization, now each species grows in abundance `greedily' until it cannot extract an energy that exceeds maintenance energy. Despite such local greedy evolution, we found that on average, the mean of the energy distribution for self-replicating agents was 100$\times$ higher than the mean for random abundances and only 10$\times$ lower than the mean for globally optimized community of machines (Fig. 4b). This result suggests that `selfish' population dynamics drives random initial abundances most of the way to abundances predicted by a global optimization algorithm, even though the `selfish' dynamics are only aware of the energy extracted by each species and do not explicitly try to maximize global energy extraction. 






Atypically large energy extraction by ecosystems was true for nearly every set of self-replicating (living) species tested (Fig. 4c), suggesting that the difference between means was not driven by only a few extremely efficient ecosystems. The population dynamics of each species naturally drove ecosystems towards such atypically large energy extraction, with $\mathcal{E}_{\text{tot}}^{\text{eco}}$ on average increasing over the dynamical trajectories of ecosystems (Fig. 4d).  

Finally, since the population dynamics of living cells (but not machines) were affected by maintenance energy $\mathcal{E}_{\text{maint}}$, we studied how the collective energy extracted depends on it.
Our simulations revealed that the total energy extracted is relatively independent of $\mathcal{E}_{\text{maint}}$ for low $\mathcal{E}_{\text{maint}}$ but falls at high $\mathcal{E}_{\text{maint}}$. At high $\mathcal{E}_{\text{maint}}$, species abundances $N^\alpha$ are the limiting factor for fluxes in the redox transformations. But as $\mathcal{E}_{\text{maint}}$ is reduced, species abundances increase and eventually, fluxes are limited by the amount of resources $R_i + O_i$ and not by species abundances; consequently, the collective energy extracted $\mathcal{E}_{\text{eco}}^{\text{tot}}$ is relatively constant.

For ecosystems of machines, collective energy extracted depends on both the energy extracted by each individual machine, as well as the number of individuals ($\mathcal{E}_{\text{eco}}^{\text{tot}} = \sum_{\alpha=1}^S \mathcal{E}_{\text{tot}}^\alpha N_\alpha$), unlike in ecosystems of self-replicators where it depends only on total biomass ($\sum_{\alpha=1}^S N_\alpha$), since each individual is constrained to extract maintenance energy ($\mathcal{E}_{\text{tot}}^\alpha = \mathcal{E}_{\text{maint}}$) at steady state. Hence, machine ecosystems may arrange species abundances to extract more energy collectively, even though the distribution of per capita energy extracted across species may be quite broad. 

Taken together, ecosystems of self-replicators with low maintanence energy $\mathcal{E}_{\text{maint}}$ capture energy comparable to ecosystems of machines with globally optimized abundances. The difference in energy extracted between self-organized and globally optimized ecosystems quantify a ``greedy gap'' of $10\times$ on average, i.e., the extent to which greedy self-replication constrains collective energy extracted in ecosystems of living cells. 


\section*{Discussion}
Here, we proposed a new theoretical framework to study self-organized energy extraction by ecosystems, which incorporates an essential aspect of metabolism thus far missing from most ecological models: organisms acquire energy through redox transformations of matter, not matter itself. Modeling resources as transformations is not only biologically accurate but also provides a modeling framework  
to address fundamental questions about thermodynamic constraints on the self-organization of ecosystems. 

Using this model, we studied the impact of closure to matter and redox metabolism on nutrient cycling in ecosystems. By sampling large random ensembles that satisfy these constraints, we found that ecosystems converge in similar thermodynamic features, such as nutrient cycling fluxes and the total energy extracted. We also found that ecosystems converge to a lesser degree when the extent of detailed balance breaking increases, i.e., when the available potential energy from light increases. We also find that the collective energy extraction is remarkably high, given that dynamics through which ecosystems assemble in our model involve local `selfish' growth rules with no awareness of the global collective energy extracted. That ecosystems extracted atypically large energy --- 100-fold closer to the theoretical maximum when compared with ecosystems with random abundances --- from their external environment (light) is an emergent feature of the dynamics of our model, arising purely from basic constraints introduced by acknowledging redox transformations and closure to matter in ecosystems.


While there is a growing body of work documenting functional convergence in ecosystems \cite{louca2016high, louca2018function, estrela2022functional, goldford2018emergent}, our work expands the domain of functional convergence to new, explicitly thermodynamic features and the impact of the environmental driving potential.
Our basic result is that functional convergence is strongest for weakly-driven near-equilibrium ecosystems; strong external driving potentials decrease convergence. This result suggests a deep theoretical connection between different manifestations of functional convergence as representing the multiplicity of ways in which communities break detailed balance. 

To calculate the energy extracted, we made the simplifying assumption that organisms extracted energy equal to the entire energy gap between their respective donors and acceptors. Realistically, a fraction of the energy gap is extracted and stored as chemical energy (e.g., ATP) while the rest is dissipated \cite{jin2003new,jin2005predicting}. This can be incorporated in our framework, by including suitable ``ATP coupling'' parameters for every donor-acceptor pair. We expect that such an extension of our model doing will generally increase niche competition between species coupling the same donor-acceptor pairs, and thus might decrease total energy extraction.


While this manuscript focused on materially closed, energy-limited ecosystems, our theoretical framework can be extended in a variety of ways. Examples include extending the framework to account for the dual role of resources, as sources of both energy and biomass, where organismal growth would depend on which of the two --- energy extraction or biomass generation --- are limiting.
Another is to extend the model to be spatially explicit, in order to study spatio-temporal pattern formation such as self-organized stratification in microbial mats and Winogradsky columns. 
Finally, our work can help identify likely signatures of life on redox towers in astrobiological contexts, e.g., by studying the self-organized adjusted potentials (like in Fig. 3a and Fig. S9).
 In all these questions, redox constraints are essential to the underlying phenomena. Thus our work opens up new lines of inquiry in redox ecology.


In addition to ecology, our framework could be used to understand the role of energy extracted in driving the evolution of living systems. Ecologists have widely argued that energy might serve as a natural fitness function during the evolution of biological communities \cite{lotka1922contribution}. However, natural selection acts on individuals and not on directly on community function. Extensions of our work can provide a framework to understand the tension between `selfish' evolution of individuals and collective energy extracted by an ecosystem. Such a framework could be useful in guiding the engineering of evolutionarily stable photosynthetic communities.


Another feature of our model is \emph{emergent} detailed balance breaking. Unlike in other models of non-equilibrium systems \cite{qian2007phosphorylation} --- where the extent of detailed balance breaking is a fixed external quantity --- in our work, we set a fixed external driving potential (e.g., that of light) in the redox tower but the amount of detailed balance breaking is determined by self-organization of the ecosystem (e.g., through species abundances and material abundances that change chemical potentials through product inhibition).  As a consequence, e.g., there is a minimal non-zero external drive below which there is no detailed balance breaking (Fig. 3b, gray region). 
In this way, our work suggests an ecology-inspired framework for studying the emergence of spontaneously self-organized non-equilibrium steady states (NESS), adding to prior work on the origin of dissipative structures inspired by Rayleigh-Benard convection cells and other physical systems \cite{kondepudi2014modern, england2013statistical, england2015dissipative}.

\section*{Methods}

{\small
\noindent
Please see SI Appendix for detailed materials and methods. Briefly, we constructed a theoretical framework for microbial ecosystems, whose constituent species extract energy through thermodynamically-constrained redox conversions of matter.  Our theory relied on fundamental physico-chemical constraints which we outlined in Results and further elaborate on in the SI Appendix. We then derived a dynamical model of ecosystem self-organization based on the constraints outlined in equations (4)-(6), resulting in  equations (7) and (8) in the main text by using thermodynamically accurate expressions for process rates (e.g., product inhibition and energy-dependent forcing). Using numerical simulations of these dynamical equations, we generated ensembles of ecosystems at steady state with a fixed physical environment, but different initial sets of species (Fig. 2), and with varying levels of detailed balance breaking $\mu_{h\nu}$ (Fig. 3). We then used numerical optimization and root-finding techniques to generate analogous ecosystems of non-living machines, by finding solutions of the constraint equations (4) and (6) which globally maximized the collective energy extracted given by equation (3) (Fig. 4).}

\section*{Acknowledgements}
{\small
We thank the Kavli Institute for Theoretical Physics, UCSB, where this project took root. We thank G. Birzu, J. Goldford, S. Maslov, A. Narla, S. Kuehn, O.X. Cordero and M. Tikhonov for discussions. This research was supported in part by the National Science Foundation under grant number NSF PHY-1748958. A.G. is supported by the Gordon and Betty Moore Foundation as a Physics of Living Systems Fellow through grant number GBMF4513.
}
\bibliography{bibliography}

\end{document}


\title{\texorpdfstring{Supplementary Text\\}{Supplement}Closed ecosystems can capture energy through self-organized nutrient cycles}
    \author{Akshit Goyal}
      \affiliation{Department of Physics, Massachusetts Insitute of Technology, Cambridge, MA 02139.}
    \author{Avi I. Flamholz}
      \affiliation{Division of Biology and Biological Engineering, California Institute of Technology, Pasadena, CA 91125.}
    \author{Alexander P. Petroff}
      \affiliation{Department of Physics, Clark University, Worcester, MA 01610.}
    \author{Arvind Murugan}%
      \affiliation{Department of Physics, University of Chicago, Chicago, IL 60637.}

\maketitle

\renewcommand{\theequation}{S\arabic{equation}}
\setcounter{equation}{0}
\renewcommand{\thefigure}{S\arabic{figure}}
\setcounter{figure}{0}

\section{Description of our mathematical model}
As briefly described in the main text, our model describes a dynamical closed ecosystem in which $S$ microbial species collectively recycle a set of environmental resources through sets of $R$ thermodynamically-constrained redox transformations. Each species in the ecosystem corresponds to a different metabolic type, depending on the subset of these $R$ transformations it can perform to maintain itself. We track the dynamics of species abundances, $N_\alpha$ and resource molecule concentrations $R_i$, and study the properties of steady states of this system. For simplicity, we assume fast transport kinetics across microbial species, so that the intracellular concentrations of molecules that each species can transform is the same as the extracellular concentration. This allows us to only track the dynamics of only one set of concentrations for each molecule. Further, we assume that species dynamics are driven only by energy requirements, without any biomass turnover. This assumption allows us to study energy extraction  at steady state, which for each surviving species must equal a prescribed maintenance energy $\mathcal{E}_{maint}$ (a tunable parameter of our model, studied in Fig. 4e). Thus, growing microbial species are limited only by energy, and we ignore the requirement of biomass building blocks. Similarly, we ignore the accumulation of organic matter by dying species in this work, and aim to study that model in greater detail in later work. We will now describe the model in greater detail, reproducing some sections from the main text description for the sake of completeness.

Each microbial species extracts energy through redox metabolic transformations (half-reactions) whose energy content is determined by electron and thermodynamic constraints (Fig. 1b). All $R$ resources correspond to a pair of transformations $O_i \leftrightarrow R_i$ between pairs of molecules $O_i,R_i$, representing the oxidized and reduced forms, respectively. A redox tower orders all resource pairs $(O_i, R_i)$ by their chemical potential $\mu_i$, from least energetically favorable $O_i \to R_i$ conversion (top) to most (bottom) (Fig. 1a). 

These chemical potentials $\mu_i$ are given by a standard state potential $\mu_i^{0}$ and due to thermodynamic product inhibition, an adjustment due to concentrations of , i.e., $\mu_i = \mu_i^{0} - \log O_i/R_i$.

Each species $\alpha$ may exploit a specific subset of these transformations specified by kinetic coefficients $e_{i\alpha} \geq 0$; e.g., if $e_{i\alpha} \neq 0$, species $\alpha$ can transform $R_i \to O_i$ with kinetic coefficient $e_{i\alpha}$, releasing each electron from transformation $R_i \to O_i$ at a potential $\mu_i$, which are then absorbed by another transformation $O_j \to R_j$ that the species participates in. The net potential difference drop experienced by the electron is the energy available to this species. In practice, the electrons are transferred by an electron carrier (e.g., NADH) in the cell that is at potential $\mu_{\text{carrier},\alpha}$ intermediate to $\mu_i$ and $\mu_j$. We assume there is only one electron carrier pool common to all redox transformations in species $\alpha$. Multiple electron carrier pools would in principle correspond to different carrier potentials, one for each carrier molecule.

We assume that detailed balance is broken across the redox tower because some transformations, say $R_j \to O_j$ are  coupled to an external energy (but not matter) source (e.g., coupling the transformation H$_2$O $\to$ O$_2$ to sunlight during photosynthesis). Consequently, the chemical potential of $R_j \to O_j$ is shifted $\mu_{j} = \mu_{-j} + \mu_{h \nu}$ where $\mu_{j}$ is the chemical potential for the reverse transformation $O_j \to R_j$ (not coupled to light). In the simulations in the main text, only one special transformation, corresponding to $j=3$ ($R_3 \to O_3$) couples to light. This is analogous to the H$_2$O to O$_2$ transformation. Importantly, the reverse transformation, $O_3 \to R_3$ (analogous to O$_2$ to H$_2$O) does not couple to light and has its adjusted potential given by the standard formula in the previous paragraph.

To specify species and resource dynamics, we first need to compute the per capita fluxes $f_{i\alpha}$ of all transformations $O_i \to R_i$ due to each species $\alpha$. Since multiple microbial species could catalyze the same transformation, and because each transformation cannot proceed faster than a certain timescale, we model $f_{i\alpha}$ using the following expression, similar to Michaelis-Menten kinetics with multiple competing enzymes (here, species): 
\begin{equation}
 f_{i \alpha} = \frac{ \cdot O_i }{ ( K_N + \sum_{\beta=1}^{S} \frac{N_\beta}{k_{i\beta}} )},
\end{equation}
where $k_{i\alpha}$ is the per capita kinetic coefficient corresponding to species $\alpha$ and molecule $i$, $N_\alpha$ is the abundance of species $\alpha$, $O_i$ is the concentration of the corresponding substrate of the transformation, and $K_N$ is a half-saturation constant, chosen arbitrarily to be $1$. From thermodynamics, $k_{i\alpha}$ is given by the force-flux relation:
\begin{equation}
 k_{i \alpha} = e_{i\alpha} \cdot (1 - e^{-\Delta \mu_{i\alpha}}) 
\end{equation}
where $\Delta \mu_{i\alpha} $ is the change in potential of an electron released by transformation $O_i \to R_i$ and captured by the electron carrier. Hence:
\begin{equation}
    \Delta \mu_{i\alpha} = \mu_{\text{\text{carrier}}, \alpha} - \Big(\mu_{i}^0 - \log\Big(\frac{O_i}{R_i}\Big)\Big),     
\end{equation}
where $\mu_i^0$ is the standard state chemical potential of the transformation $O_i \to R_i$ and $\mu_{\text{\text{carrier}}, \alpha}$ is the chemical potential of the electron carrier for in species $\alpha$. Here, $k_{i\alpha}$ is the net forward rate of the transformation ($J_+/J_-$ in thermodynamics), and $\Delta \mu_{i \alpha}$ is the corresponding energy difference. $e_{i\alpha}$ is a kinetic constant that specifies the per capita per unit concentration rate at which species $\alpha$ can perform transformation $i$, as described above (0 implying $\alpha$ cannot perform the transformation).

Together, all fluxes $f_{i \alpha}$ change the concentrations of $O_i, R_i$ through the following dynamics:
\begin{equation}
    \frac{dO_i}{dt} = -\sum_{\alpha=1}^{S} f_{i\alpha} N_\alpha + \sum_{\beta=1}^{S} f_{i\beta} N_\beta 
\end{equation}
where $f_{i\alpha}$ is the flux of the transformation $O_i \rightarrow R_i$ performed by an individual of species $\alpha$ (as given in equation (1)), and $f_{i\beta}$ is the flux of the transformation $R_i \to O_i$ (similar to equation (1), but proportional to the reactant concentration $R_i$, not $O_i$ by individuals of species $\beta$. The first sum goes over all species transforming $O_i \to R_i$ and the second sum over species capable of the reverse. Similar equations hold for $R_i$.

Each species $\alpha$ extracts energy with flux $\mathcal{E}_{\text{tot}}^{\alpha}$ by coupling electrons between transformations at different potentials:
\begin{eqnarray}
     \mathcal{E}_{\text{tot}}^{\alpha} &=& \sum_{i=1}^{2R} \mathcal{E}_{i\alpha} = 
     \sum_{i=1}^{2R} f_{i\alpha} \cdot \Delta \mu_{i\alpha}. 
\end{eqnarray}
A species grows in abundance if this captured energy exceeds a prescribed per capita maintenance energy $\mathcal{E}_{\text{maint}}$:

\begin{equation}
    \frac{1}{N_\alpha} \frac{dN_\alpha}{dt} = \mathcal{E}_{\text{tot}}^{\alpha} - \mathcal{E}_{\text{maint}}.
\end{equation}

%

Finally, to balance all cycles at steady state, i.e., to conserve matter, as species in the ecosystem couple different half-reactions and transform resources from one form to another, all resource cycles must be balanced (Fig. 1e). Together, these constraints can be summarized as:

\begin{align}
    \underline{\text{electron conservation:}} && \sum_{i=1}^{2R} f_{i\alpha} &= 0,\\
    \underline{\text{energy requirement:}} && \sum_{i=1}^{2R} f_{i\alpha} \Delta \mu_{i\alpha} &= \mathcal{E}_{\text{maint}},\\
    \underline{\text{matter conservation:}} && \sum_{\alpha=1}^{S} N_\alpha f_{i\alpha} &= 0.
\end{align}

The last equation amounts to assuming that the ecosystem is fully closed to matter, and open only to an external source of energy (here, light energy $\mu_{h\nu}$) that breaks detailed balance for the chemical potentials $\mu_i$. While we use a fully materially closed ecosystem as an extreme case to illustrate our model, our key results hold for partially closed ecosystems as well (Fig. S3), where some of the resources can be exchanged with the environment and equation (S4) is modified as follows: 

\begin{equation}
    \frac{dO_i}{dt} = -\sum_{\alpha=1}^{S} f_{i\alpha} N_\alpha + \sum_{\beta=1}^{S} f_{i\beta} N_\beta + \kappa_i - \delta_i O_i,
\end{equation}
where $\kappa_i$ specifies the rate of influx of molecule $O_i$, and $\delta_i$ specifies its specific (per unit concentration) dilution rate.

\section{Complete set of dynamical equations\\ for a 2-species, 2 cycle system}
For concreteness, we now explicitly write down the set of dynamical equations for the simplest possible ecosystem in our model: a 2-species system (e.g., heterotroph-phototroph) with 2 resource cycles (e.g., H$_2$O/O$_2$ and CO$_2$/org C). This would be an ecosystem where $S=2$ and $R=2$, as opposed to ecosystems with $R=3$ and $S$ between 3 and 10 that we study in the main text. We will assume that the $O_1/R_1$ pair corresponds to CO$_2$/org C and $O_2/R_2$ pair corresponds to O$_2$/H$_2$O respectively. 

We will represent the phototroph --- which transforms $R_2 \to O_2$ and $O_1 \to R_1$ ---  with label $P$ and abundance $N_P$, and heterotroph --- which transforms $O_2 \to R_2$ and $R_1 \to O_1$ --- with label $H$ and abundance $N_H$, respectively. Analogous to photosynthesis, only the $R_2 \to O_2$ transformation will coupled to light. Following the previous section, the population dynamics for the two species are given by the following equations:

\begin{align}
    \frac{1}{N_P} \frac{dN_P}{dt} &= \frac{R_2}{K_N + \frac{N_P}{e_{2P} (1-\exp(-\Delta \mu_{2P}))}} (\Delta \mu_{2P} - \mu_{h\nu}) + \frac{O_1}{ K_N + \frac{N_P}{e_{1P} (1-\exp(-\Delta \mu_{1P}))}} \Delta \mu_{1P} - \mathcal{E}_{maint}, \\
    \frac{1}{N_H} \frac{dN_H}{dt} &= \frac{O_2}{K_N + \frac{N_H}{e_{2H} (1-\exp(-\Delta \mu_{2H}))}} \Delta \mu_{2H} + \frac{R_1}{ K_N + \frac{N_H}{e_{1H} (1-\exp(-\Delta \mu_{1H}))}} \Delta \mu_{1H} - \mathcal{E}_{maint},
\end{align}
where:
\begin{equation}
    \Delta \mu_{i\alpha} = \mu_{\text{\text{carrier}}, \alpha} - \Big(\mu_{i}^0 - \log\Big(\frac{O_i}{R_i}\Big)\Big)
\end{equation} 
for all $i\in{1,2}$ and $\alpha\in{P,H}$. The first term in equation (S11) has an additional term $\mu_{h_\nu}$ that indicates that the transformation $R_2 \to O_2$ is coupled to light and breaks detailed balance. The dynamics for the $\mu_{\text{\text{carrier}}, \alpha}$ are given by the following equations:

\begin{align}
    \frac{d}{dt} \mu_{\text{\text{carrier}}, P} &= \frac{R_2}{K_N + \frac{N_P}{e_{2P} (1-\exp(-\Delta \mu_{2P}))}} - \frac{O_1}{ K_N + \frac{N_P}{e_{1P} (1-\exp(-\Delta \mu_{1P}))}}, \\
    \frac{d}{dt} \mu_{\text{\text{carrier}}, H} &= \frac{O_2}{K_N + \frac{N_H}{e_{2H} (1-\exp(-\Delta \mu_{2H}))}} - \frac{R_1}{ K_N + \frac{N_H}{e_{1H} (1-\exp(-\Delta \mu_{1H}))}}.
\end{align}

The nutrient dynamics are given by the following dynamical equations:

\begin{align}
    \frac{dR_1}{dt} &= -\frac{R_1}{ K_N + \frac{N_H}{e_{1H} (1-\exp(-\Delta \mu_{1H}))}} N_H + \frac{O_1}{ K_N + \frac{N_P}{e_{1P} (1-\exp(-\Delta \mu_{1P}))}} N_P,\\
    \frac{dO_1}{dt} &= \frac{R_1}{ K_N + \frac{N_H}{e_{1H} (1-\exp(-\Delta \mu_{1H}))}} N_H - \frac{O_1}{ K_N + \frac{N_P}{e_{1P} (1-\exp(-\Delta \mu_{1P}))}} N_P,\\
    \frac{dR_2}{dt} &= -\frac{R_2}{K_N + \frac{N_P}{e_{2P} (1-\exp(-\Delta \mu_{2P}))}} N_P + \frac{O_2}{K_N + \frac{N_H}{e_{2H} (1-\exp(-\Delta \mu_{2H}))}} N_H,\\
    \frac{dR_2}{dt} &= +\frac{R_2}{K_N + \frac{N_P}{e_{2P} (1-\exp(-\Delta \mu_{2P}))}} N_P - \frac{O_2}{K_N + \frac{N_H}{e_{2H} (1-\exp(-\Delta \mu_{2H}))}} N_H.
\end{align}

Examples of dynamics from this system are shown in Fig. S1. Similar examples from more complex many species ecosystems with $R=3$ are shown in Fig. S2.

\section{Parameters and initial conditions used for simulations}
\begin{table}[h]
\footnotesize
\centering
\renewcommand{\arraystretch}{1.3}
\setlength{\tabcolsep}{15pt}
\begin{tabular}{c c c}
\hline
Symbol & Interpretation & Typical value (if applicable)\\
\hline
$R$ & Number of resource cycles & 3 \\
$S$ & Number of species added & 3--10 \\
$N_{pool}$ & Number of species in the random ensemble & 1,000 \\
$K_N$ & Half-saturation coefficient for flux & 1 \\
$O_i + R_i$ & Total amount for resources in each cycle  & 1 \\
$e_{i\alpha}$ & Kinetic coefficient for transformation $i$ by species $\alpha$ & $\mathcal{U}(0,1)$ \\
$\mu_{i}^0$ & Standard state potentials of redox pairs &  $\{+0.8, -0.4, -0.1\}$ \\
$\mu_{h\nu}$ & Potential of external driving force (light)  & 2 \\
\hline
\end{tabular}
\end{table}

For each simulation, we assumed random initial conditions for all species abundances and electron carrier potentials, such that they were uniformly distributed between 0 and 1. For each resource cycle $i$, we uniformly partitioned the total resource amount $O_i + R_i$ at random. We ran all simulations using the Radau solver for 1,000 time steps. We simulated equations for species abundances, resource concentrations and electron carrier potentials to steady state (equations (S7)--(S9)).

As $O_i,R_i$ concentrations and their corresponding chemical potentials changed, each species reorganized its internal fluxes to balance the net flow of electrons and subsequently captured a different energy flux from the environment. Some species continued to capture more than $\mathcal{E}_{\text{\text{maint}}}$ and grow, while others captured less than $\mathcal{E}_{\text{maint}}$ and decreased in abundance, some even going extinct. This feedback between species and resources continued until the ecosystem self-organized into a steady state where individuals of every surviving species captured energy flux $\mathcal{E}_{\text{maint}}$ with a balanced internal redox electron flux, and all resource cycles were balanced at certain fluxes $\phi_i$. Together, all surviving individuals in the ecosystem captured a total energy flux $\mathcal{E}_{\text{tot}}^{\text{eco}} = \sum_{\alpha=1}^{S} \mathcal{E}_{\text{tot}}^\alpha N_\alpha$ (Fig. 2a). By simulating 1,000 such ecosystems, we sampled a large space of ecosystems that obeyed thermodynamic constraints arising from using redox transformations as resources.

\section{Measuring the volumes of ecosystem solution spaces}
To compare the volumes of the space of possible ecosystem solutions --- both in terms of species abundances and nutrient cycle fluxes --- simulated using our model, we projected all solutions to a common, two-dimensional space. This projection was necessary for any comparison because the species and flux spaces have different inherent dimensionalities ($S=N_{pool}$ and $R=3$ respectively). To preserve pairwise Euclidean distances between ecosystems and allow for a fair comparison of projected volumes, we first normalized all Euclidean distances in species and flux spaces by the maximum Euclidean distance, and then used multi dimensional scaling (MDS), for which standard algorithms exist. We projected all species abundance and cycle flux vectors to a common two-dimensional space (Fig. 2d). To estimate the volume of each space, we measured the `volume' (area) of the convex hull of the projected set of species and flux points respectively. We then repeated this procedure for each value of $S$, each of which we had an ensemble of 1,000 ecosystems (points) for. We then normalized all measured volumes by the the largest volume (so that the largest volume was 1), and plotted this for the species (green) and fluxes (red) solutions in Fig. 2e.

\section{Building ensembles of machine ecosystems}
To build ensembles of ecosystems of machines, we performed constrained optimization for equations (S7) and (S9), instead of dynamically evolving the equations (S4), (S6) and equations of the form (S14-15). Specifically, we used the same set of parameters as for biological ecosystems, but for each machine ecosystem with randomized abundances, we numerically found a solution to the constraints (S7) and (S9) corresponding to them for a randomized set of species abundances, chosen uniformly in log-space between $10^{-5}$ and $1$. To find each solution, we used  standard non-linear least squares methods implemented in SciPy. For ecosystems of machines with optimized abundances for the same parameters, we chose species abundances so that they  maximized the total energy extraction $\mathcal{E}_{\text{tot}}^{\text{eco}} = \sum_{\alpha=1}^{S} \mathcal{E}_{\text{tot}}^\alpha N_\alpha$, while subject to the constraints (S7) and (S9) corresponding to the parameters. For this, we used the SciPy constrained optimization routine `minimize' with method SLSQP.

\clearpage
\section{Supplementary Figures}

\vfill
\begin{figure}[h]
    \centering
    \includegraphics[scale=0.5]{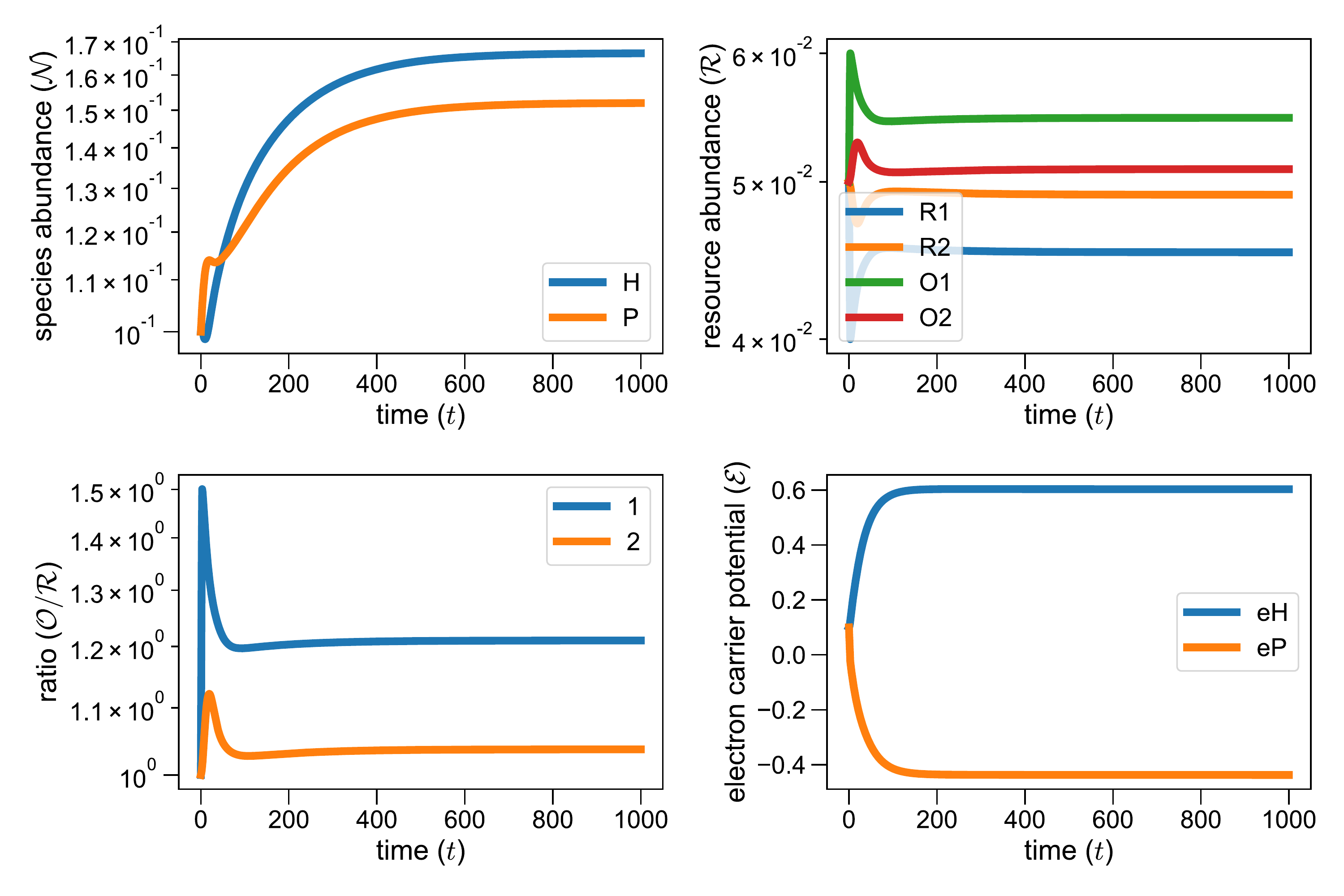}
    \caption{\textbf{Example of dynamics in a 2-species, 2-resource cycle ecosystem.} Example of dynamics of an ecosystem with $S=2$ and $R=2$, corresponding to the phototroph-heterotroph system outlined in section II, with default parameters and random initial conditions (section III). The top-left panel shows the species dynamics ($H$ representing the heterotroph in blue and $P$, the phototroph in orange). The top-right panel shows the resource dynamics (legend shown), while the bottom panels show the dynamics of resource ratios (left) and electron carrier potentials in both species (right).}
    \label{figS1}
\end{figure}
\vfill

\clearpage
.
\vfill
\begin{figure}[h]
    \centering
    \includegraphics[scale=0.5]{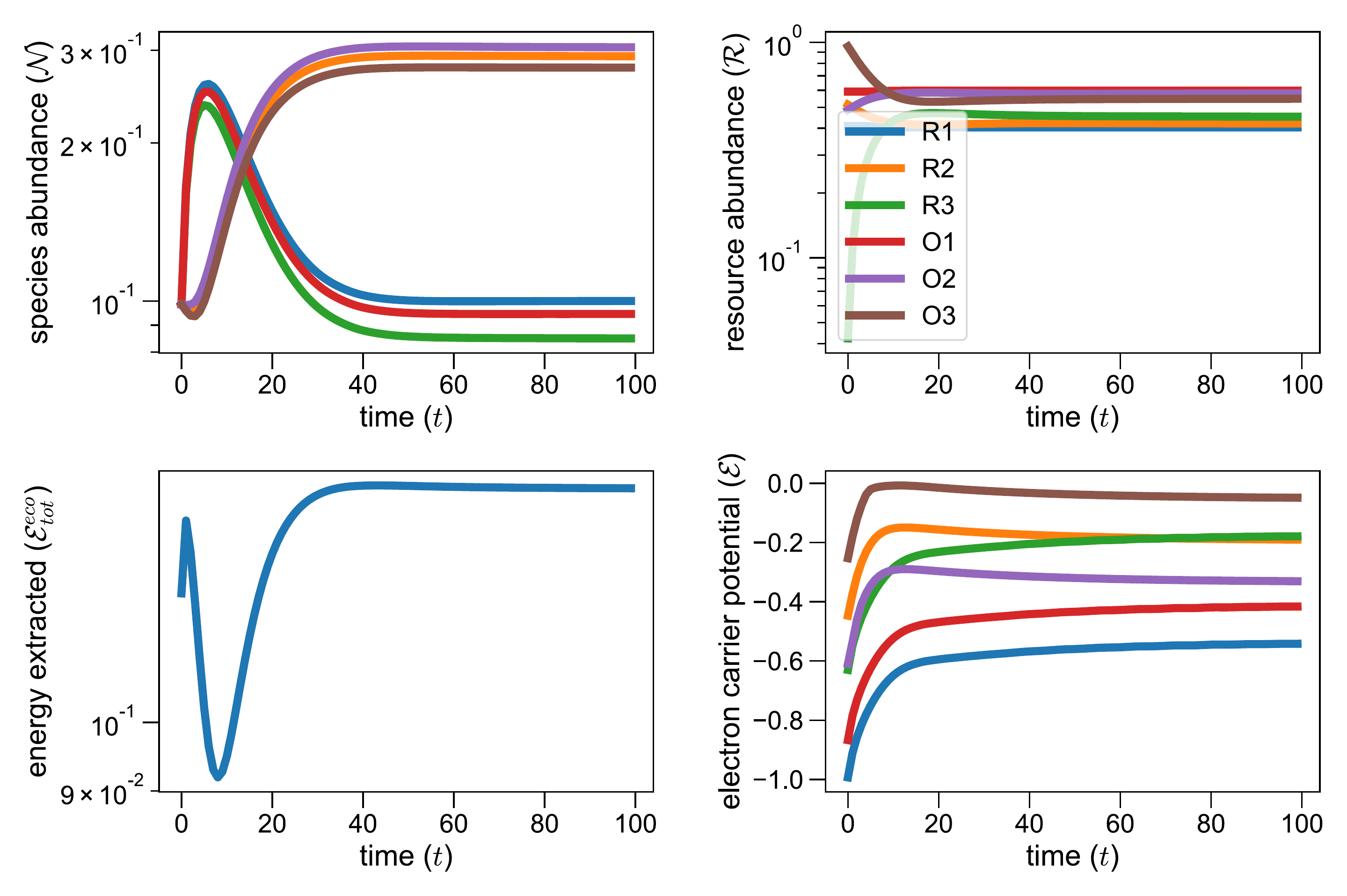}
    \caption{\textbf{Example of dynamics in a 6-species, 3-resource cycle ecosystem.} Example of dynamics of an ecosystem with $S=6$ and $R=3$, corresponding to a complex ecosystem outlined in the main text, Fig. 2, with default parameters and random initial conditions (section III). The top-left panel shows the species dynamics (each color representing a different random species). The top-right panel shows the resource dynamics (legend shown), while the bottom panels show the dynamics of the total energy extracted by the ecosystem (left) and electron carrier potentials in all species (right).}
    \label{figS2}
\end{figure}
\vfill

\clearpage
.
\vfill
\begin{figure}[h]
    \centering
    \includegraphics[scale=0.5]{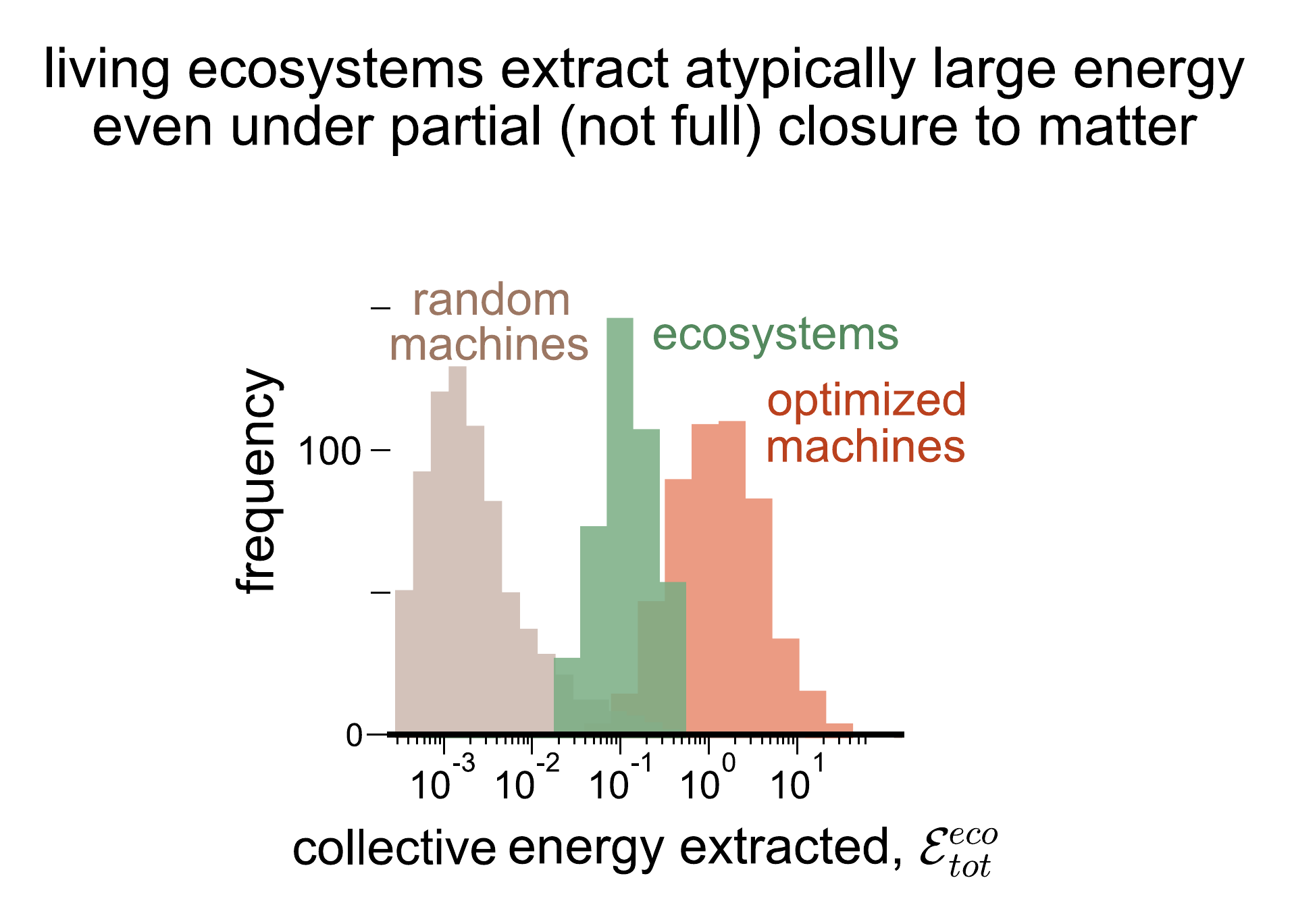}
    \caption{\textbf{Our main result is robust even if ecosystems are only partial closed to matter.}  Histograms of the total energy extracted $\mathcal{E}_{\text{tot}}^{\text{eco}}$ by ecosystems of machines with random abundances (gray; representing initial conditions of ecosystem assembly), ecosystems of self-replicators with abundances self-organized by birth-death dynamics based on maintenance energy (green), and machines with abundances chosen to maximize $\mathcal{E}_{\text{tot}}^{\text{eco}}$ (orange). Similar to Fig. 4b, but simulated using one of the 3 resource cycles ($O2  \leftrightarrow R_2$) open to matter ($\kappa_2 = 0.1, \delta_2 = 0.1$). On average, ecosystems still extract energy closer to optimal than those comprising machines with randomized abundances. 
    }
    \label{figS3}
\end{figure}
\vfill
\clearpage

.
\vfill
\begin{figure}[h]
    \centering
    \includegraphics[scale=0.7]{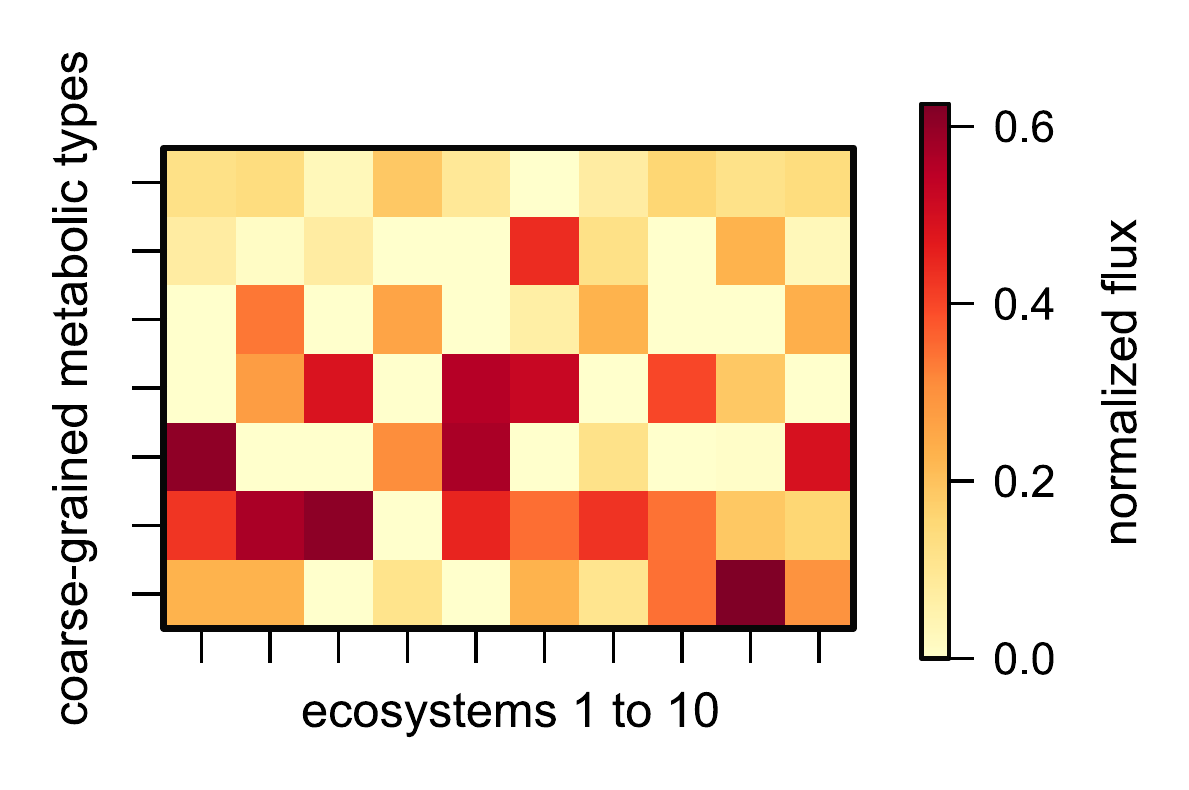}
    \caption{\textbf{Ecosystem structure is more convergent when species are coarse-grained by metabolic types.} Heatmaps showing examples from 10 of the 1,000 randomly assembled ecosystems in Fig. 2, showing the combined abundances of species when grouped by their ``metabolic type'', i.e., by the subset of transformations they can perform with $e_{i\alpha} \neq 0$. Each row shows a metabolic type, while each column shows an example ecosystem.
    }
    \label{figS4}
\end{figure}
\vfill

\clearpage
.
\vfill
\begin{figure}[h]
    \centering
    \includegraphics[scale=0.5]{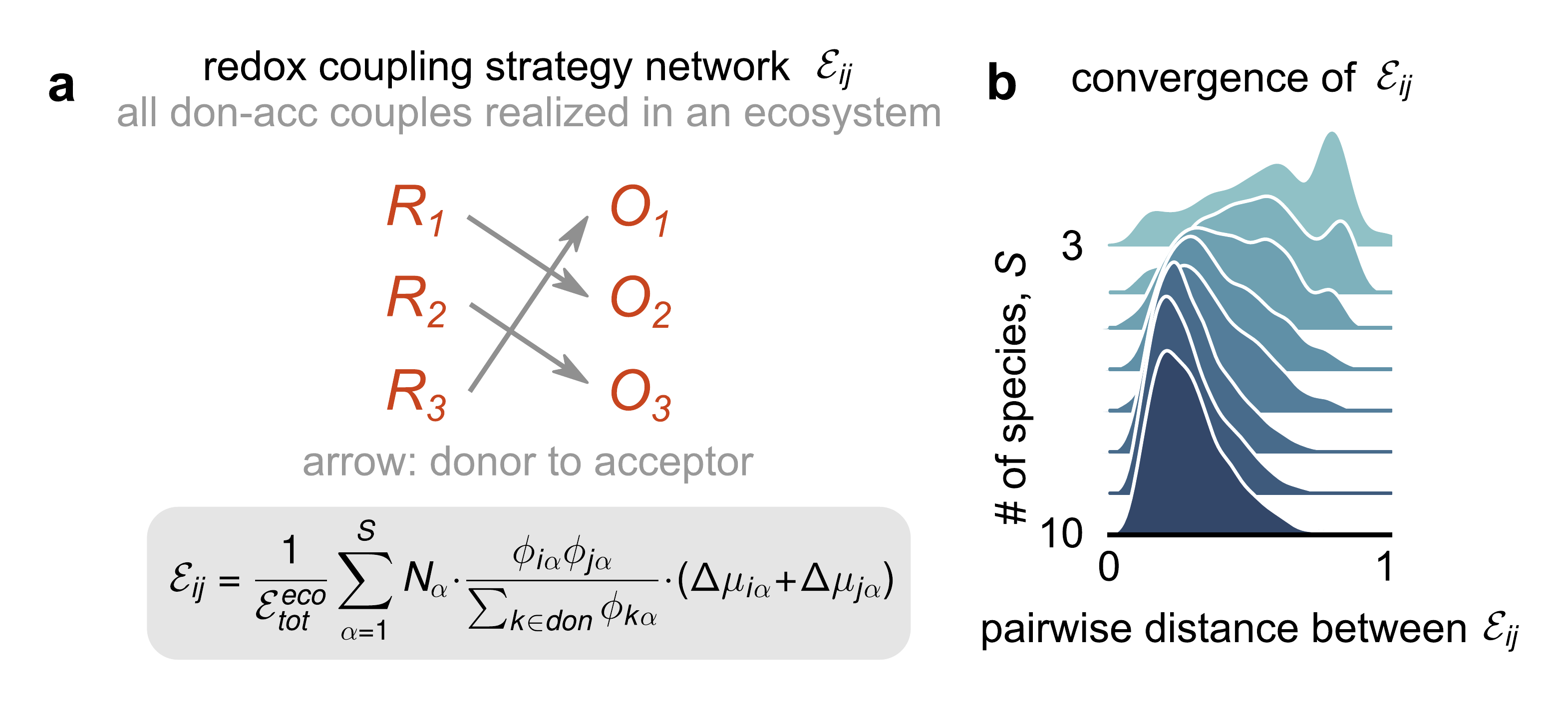}
    \caption{\textbf{Global redox ``strategies'' converge with increasing species diversity.} (a) Schematic showing an example of a global redox strategy for an entire ecosystem, which quantifies the fraction $\mathcal{E}_{ij}$ of the total ecosystem energy flux $\mathcal{E}_{tot}^{eco}$ that is obtained by coupling transformations $i$ and $j$. We compute $\mathcal{E}_{ij}$ as shown in the gray box, where $\phi_{i\alpha} = f_{i\alpha} N_\alpha$ is the contribution to the total flux in resource $i$ by all individuals of species $\alpha$. Notably, $\sum_{i\neq j} \mathcal{E}_{ij} = 1$ is a matrix that is normalized by definition. (b) Histograms of the pairwise Euclidean distance between the global redox strategies $\mathcal{E}_{ij}$, as a function of number of species added $S$, for all ecosystems simulated in Fig. 2. The average distance (and its variance) decreases with increasing diversity, suggesting that the strategies implemented across different ecosystems become more similar (converge).
    }
    \label{figS5}
\end{figure}
\vfill

\begin{figure}[h]
    \centering
    \includegraphics[scale=0.5]{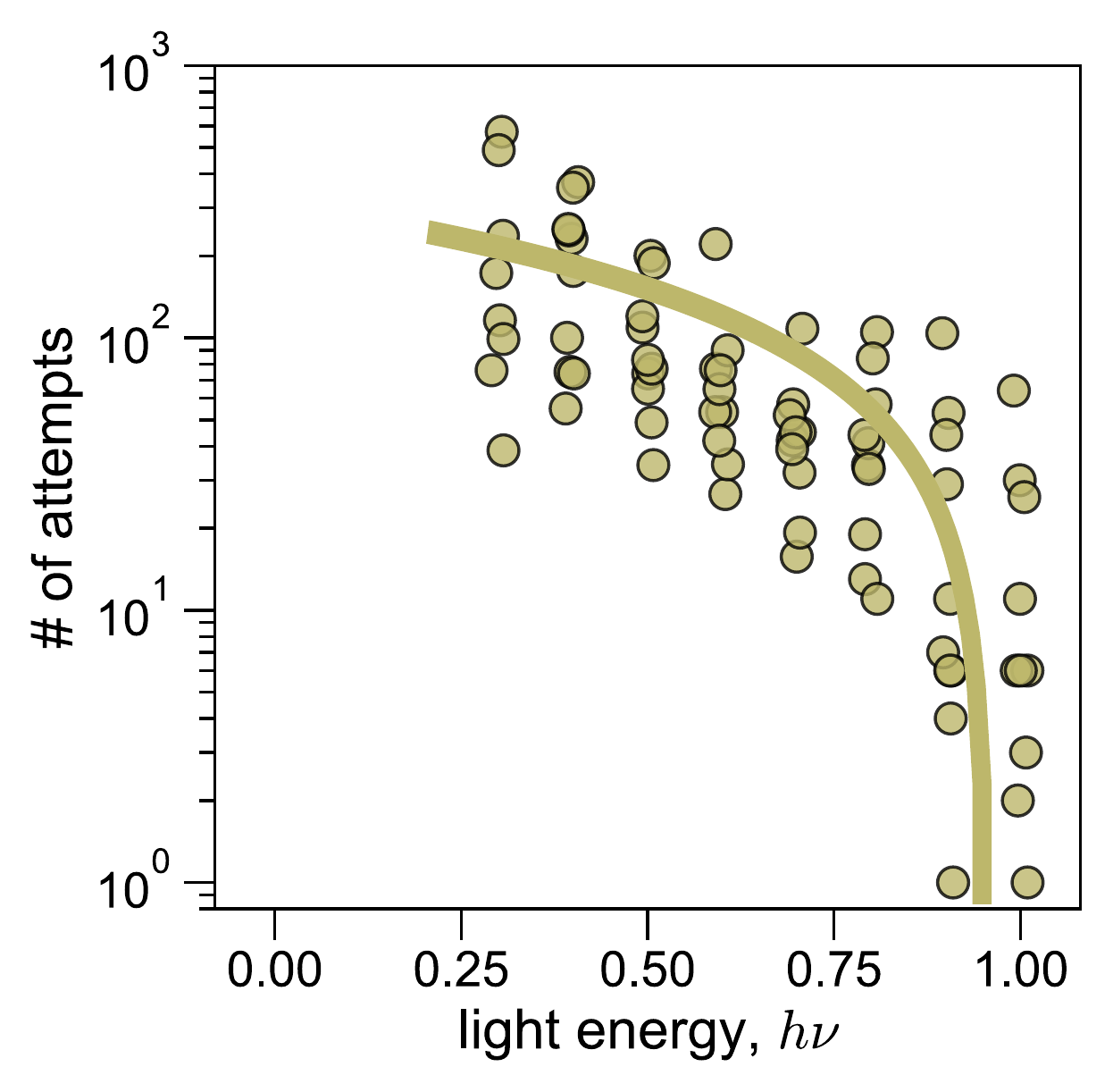}
    \caption{\textbf{Once enough driving light energy is available, randomly assembled ecosystems rarely collapse.} Scatter plot showing the number of simulations with random initial conditions (section III) that we need to run before we arrive at an ecosystem that can self-organize to capture nonzero energy, i.e., that does not collapse. We show the number of attempts as a function of the light energy driving the ecosystems $\mu_{h\nu}$, as explained in section I. Ecosystems cannot self-organize below a minimum $\mu_{h\nu}$ as in Fig. 3. Once the driving energy is large enough, ecosystems can almost always self-organize (the number of attempts become close to 1).}
    \label{figS6}
\end{figure}

\begin{figure}[h]
    \centering\
    \includegraphics[scale=0.5]{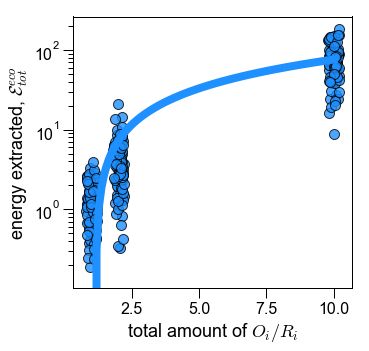}
    \caption{\textbf{Once there is sufficiently large light energy, the total amount of resources limit the energy extracted by ecosystems.} Scatter plot showing the total energy extracted $\mathcal{E}_{tot}^{eco}$ as a function of the total amount of resources  $\sum(O_i + R_i)$ supplied to closed ecosystems. Each point represents a randomly assembled ecosystem from a large ensemble of species, as in Figs. 2 and 3, but assembled with different total resource amounts.
    }
    \label{figS7}
\end{figure}

\begin{figure}[h]
    \centering
    \includegraphics[scale=0.55]{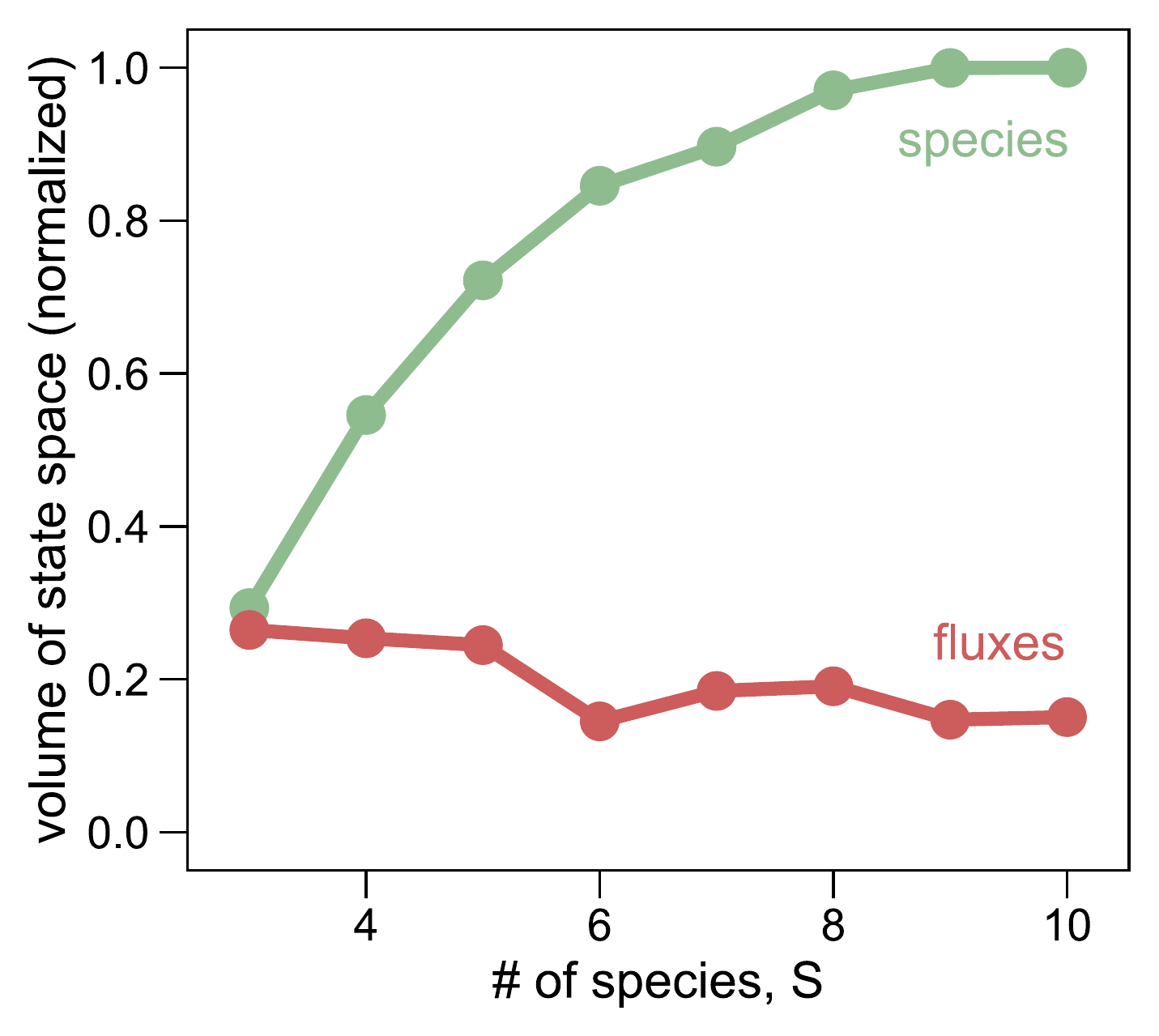}
    \caption{\textbf{Our results about flux convergence are robust to the addition of more resource cycles.}  Line plot, similar to Fig. 2e, but with $R=5$ instead of $3$. The plot shows how the volume of the species (green) and flux (red) spaces scales with the number of species added, $S$, in assembled ecosystems. As in Fig. 2e, the flux space volume grows much slower than species space volume, indicating convergence in the function (fluxes) of self-organized ecosystems.}
    \label{figS8}
\end{figure}

\begin{figure}[h]
    \centering
    \includegraphics[scale=0.55]{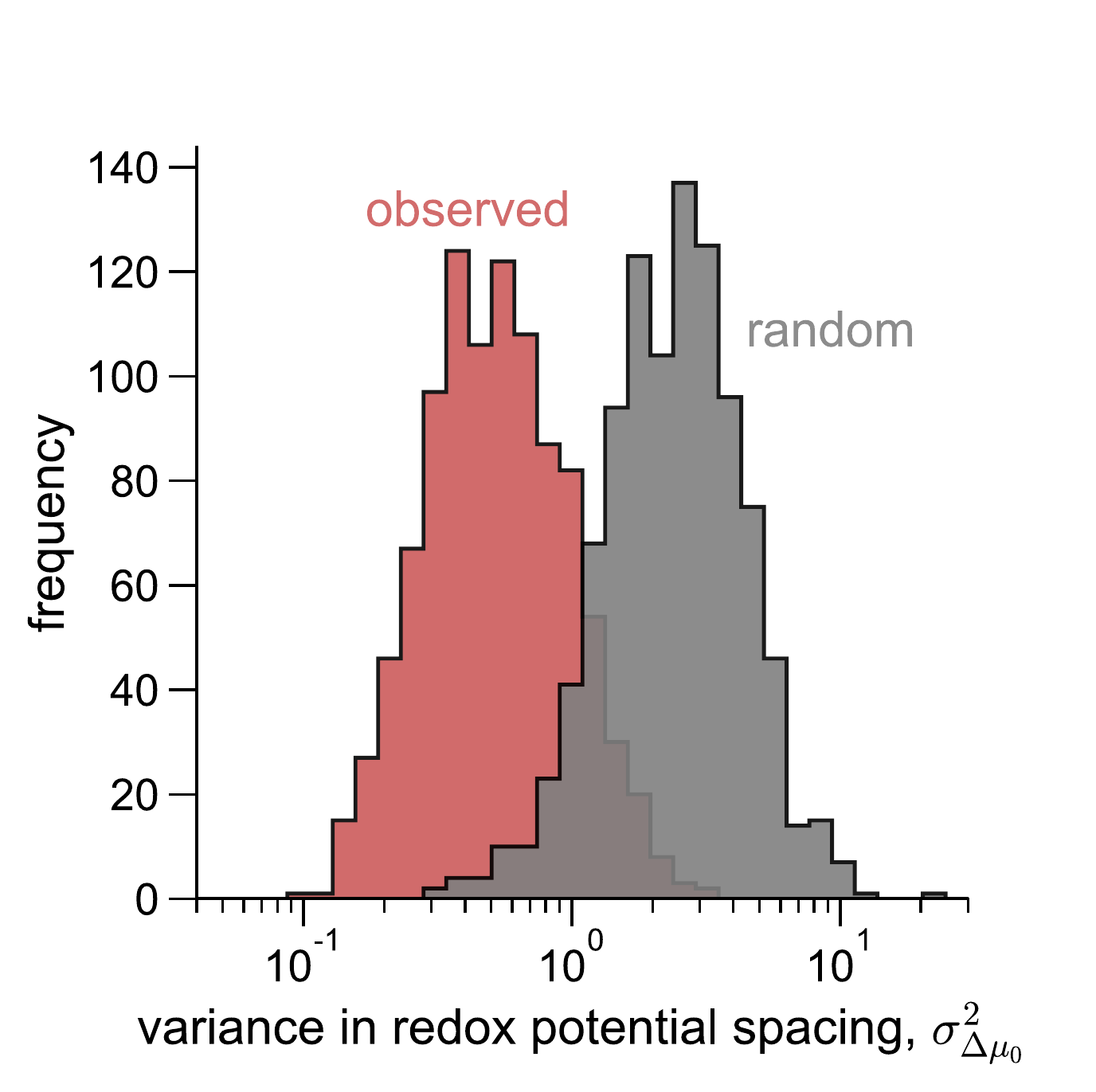}
    \caption{\textbf{Self-organized redox potentials are more equally spaced than expected by chance.} Histograms of the variance in spacing between adjusted redox potentials for ecosystems at steady state (red), compared with the variance in spacing when the potentials are randomized, i.e., spread uniformly in the same range. The observed potentials show roughly 10-fold lower variance in spacing.  
    }
    \label{figS9}
\end{figure}